\documentclass[journal=jpcbfk,manuscript=article]{achemso}
\usepackage{longtable}
\usepackage{multirow}
\usepackage{amsmath}
\usepackage{subfigure}
\usepackage[usenames,dvipsnames]{color}
\usepackage{soul}
\usepackage{threeparttable}
\usepackage{xr}
\usepackage{graphicx}
\usepackage{amsmath,amssymb}
\usepackage{caption}
\usepackage{color}
\usepackage{dcolumn}
\usepackage{bm}
\usepackage{float}
\usepackage{subfig}
\usepackage{textcomp}
\usepackage{url}
\usepackage[normalem]{ulem}

\makeatletter
\newcommand*{\addFileDependency}[1]{
  \typeout{(#1)}
  \@addtofilelist{#1}
  \IfFileExists{#1}{}{\typeout{No file #1.}}
}
\makeatother
 
\newcommand*{\myexternaldocument}[1]{%
    \externaldocument{#1}%
    \addFileDependency{#1.tex}%
    \addFileDependency{#1.aux}%
}
\myexternaldocument{si}

\author{Seyedeh Maryam Salehi} \affiliation[University of
  Basel]{Department of Chemistry, University of Basel,
  Klingelbergstrasse 80 , CH-4056 Basel, Switzerland.}
\author{Markus Meuwly} \affiliation[University of Basel]{Department of
  Chemistry, University of Basel, Klingelbergstrasse 80 , CH-4056
  Basel, Switzerland.}  \email{m.meuwly@unibas.ch}

\title {Site-Selective Dynamics of Azidolysozyme}

\begin{document}

\begin{abstract}
The spectroscopic response of and structural dynamics around all
azido-modified alanine residues (AlaN$_3$) in Lysozyme is
characterized. It is found that AlaN$_3$ is a positionally sensitive
probe for the local dynamics, covering a frequency range of $\sim 15$
cm$^{-1}$ for the center frequency of the line shape. This is
consistent with findings from selective replacements of amino acids in
PDZ2 which reported a frequency span of $\sim 10$ cm$^{-1}$ for
replacements of Val, Ala, or Glu by azidohomoalanine (AHA). For the
frequency fluctuation correlation functions (FFCFs) the long-time
decay constants $\tau_2$ range from $\sim 1$ to $\sim 10$ ps which
compares with experimentally measured correlation times of 3
ps. Attaching azide to alanine residues can yield dynamics that decays
to zero on the few ps time scale (i.e. static component $\Delta_0 \sim
0$ ps$^{-1}$) or to a remaining, static contribution of $\sim 0.5$
ps$^{-1}$ (corresponding to 2.5 cm$^{-1}$), depending on the local
environment on the 10 ps time scale. The magnitude of the static
component correlates qualitatively with the degree of hydration of the
spectroscopic probe. Although attaching azide to alanine residues is
found to be structurally minimally invasive with respect to the
overall protein structure, analysis of the local hydrophobicity
indicates that the hydration around the modification site differs for
modified and unmodified alanine residues, respectively.
\end{abstract}

\section{Introduction}
Understanding the structural and functional dynamics of proteins in
the condensed phase is a prerequisite for characterizing cellular
processes at a molecular level.\cite{schuler:2017} As an example,
knowledge of the mechanisms and physical principles underlying
protein-ligand recognition facilitates rational drug design for
treatment of diseases.\cite{zhou:2016,zhang:2019} One possibility to
directly and quantitatively probe the structure and dynamics of
proteins and protein-ligand complexes is vibrational, in particular
2-dimensional infrared (2D-IR) spectroscopy.\cite{2DIRbook-Hamm-2011}
\\

\noindent
Given the spectroscopic response of proteins in solution that cover
the range up to $\sim 1700$ cm$^{-1}$ and frequencies above $\sim
2800$ cm$^{-1}$, suitable vibrational labels should absorb in the
window between $\sim 1700$ and $\sim 2800$
cm$^{-1}$.\cite{gai:2011,hamm.rev:2015} A range of such probes has
been proposed and considered in the past, including
cyanophenylalanine\cite{thielges:2015}, nitrile-derivatized amino
acids,\cite{gai:2003} the sulfhydryl band of
cysteines,\cite{hamm:2008} deuterated carbons,\cite{romesberg:2011}
non-natural labels consisting of metal-tricarbonyl modified with a
-(CH$_2$)$_{\rm n}$- linker,\cite{zanni:2013} nitrile
labels,\cite{fayer.ribo:2012} cyano\cite{romesberg.cn:2011} and
SCN\cite{bredenbeck:2014} groups, or cyanamide.\cite{cho:2018} Another
promising and sensitive label that was recently used is
azidohomoalanine (AHA)\cite{hamm:2012} for which it has been
demonstrated that it can be used to characterize the recognition site
between the PDZ2 domain and its binding partner to provide
site-specific insight into the underlying mechanisms of how signaling
proteins function.\cite{stock:2018}\\

\noindent
The noncanonical amino acid AHA absorbs around $\sim 2100$ cm$^{-1}$
with a comparatively large extinction coefficient of up to 400
M$^{-1}$cm$^{-1}$.\cite{hamm:2012} From a preparative perspective
attachment of N$_3^{-}$ to alanine (to give AlaN$_3$) and AHA and
incorporation at almost any position of a protein through known
expression techniques has been demonstrated.\cite{bertozzi:2002}
Furthermore, attachment of an N$_3^-$ probe is a spatially small
modification and the chemical perturbations induced are expected to be
small. This makes AlaN$_3$ and AHA worthwhile modifications to probe
local protein dynamics.\\

\noindent
Optical spectroscopy, and especially two-dimensional infrared (2D-IR)
spectroscopy, quantitatively provides information about the structure
and dynamics of the solvent environment surrounding a probe
molecule.\cite{hamm:2015} Such techniques can also be used to measure
the subpicosecond to picosecond dynamics in condensed-phase
systems. With that, the coupling between inter- and intramolecular
degrees of freedom such as the hydrogen bonding network in solution,
or structural features of biological macromolecules can be
investigated by monitoring the fluctuation of fundamental vibrational
frequencies of a probe molecule or ligand attached to a complex or a
biological macromolecule. The possibility to use infrared spectroscopy
for characterizing protein-ligand complexes has already been proposed
for the nitrile containing inhibitor IDD743 complexed with WT and
mutant human aldose reductase\cite{Suydam.halr.sci.2006} and
explicitly demonstrated for cyano-benzene in the active site of WT and
mutant lysozyme.\cite{MM.lys:2017} \\

\noindent
The AHA label was previously used in 2D-IR spectroscopy studies of
ligand binding to the PDZ2 domain.\cite{hamm:2012} The spectral
changes observed for various modified pepdidic binders were consistent
with the known X-ray structure of the wild-type peptide bound to the
protein. This suggests that AHA is suitable as a specific IR reporter
and to highlight subtle changes of the electrostatic environment on
the protein surface.\cite{stock:2018} In the present work, attaching
N$_3^-$ to all alanine residues in Lysozyme in succession is used to
characterize the local dynamics around such modification site.\\

\noindent
Recent investigations have demonstrated that the vibrational dynamics
of N$_3^{-}$ in the gas phase and in solution can be captured
quantitatively.\cite{Salehi:N3-JPCB2019} Based on high-level
electronic structure calculations at the multi-reference configuration
interaction (MRCI) level of theory and representing the 3-dimensional
potential energy surface (PES) as a reproducing kernel Hilbert space
(RKHS),\cite{RKHS-Rabitz-1996,MM.rkhs:2017} the infrared spectroscopy
in the gas and condensed phase was correctly described. Also, the
frequency correlation function exhibited time scales consistent with
experiment which suggests that the coupling between solvent and solute
was correctly described.\\

\noindent
The present work explores the local dynamics of all alanine residues
in lysozyme as a typical model system by attaching N$_3^{-}$ as a
spectroscopic reporter. First, the computational methods are
summarized. Then, the structural dynamics and spectroscopy for all 14
AlaN$_3$ labels is discussed and the local dynamics and hydration are
explored. Finally, conclusions are drawn.\\

\section{Methods}

\subsection{Molecular Dynamics Simulations} 
\begin{figure}[H]
\begin{center}
\includegraphics[width=0.8\textwidth]{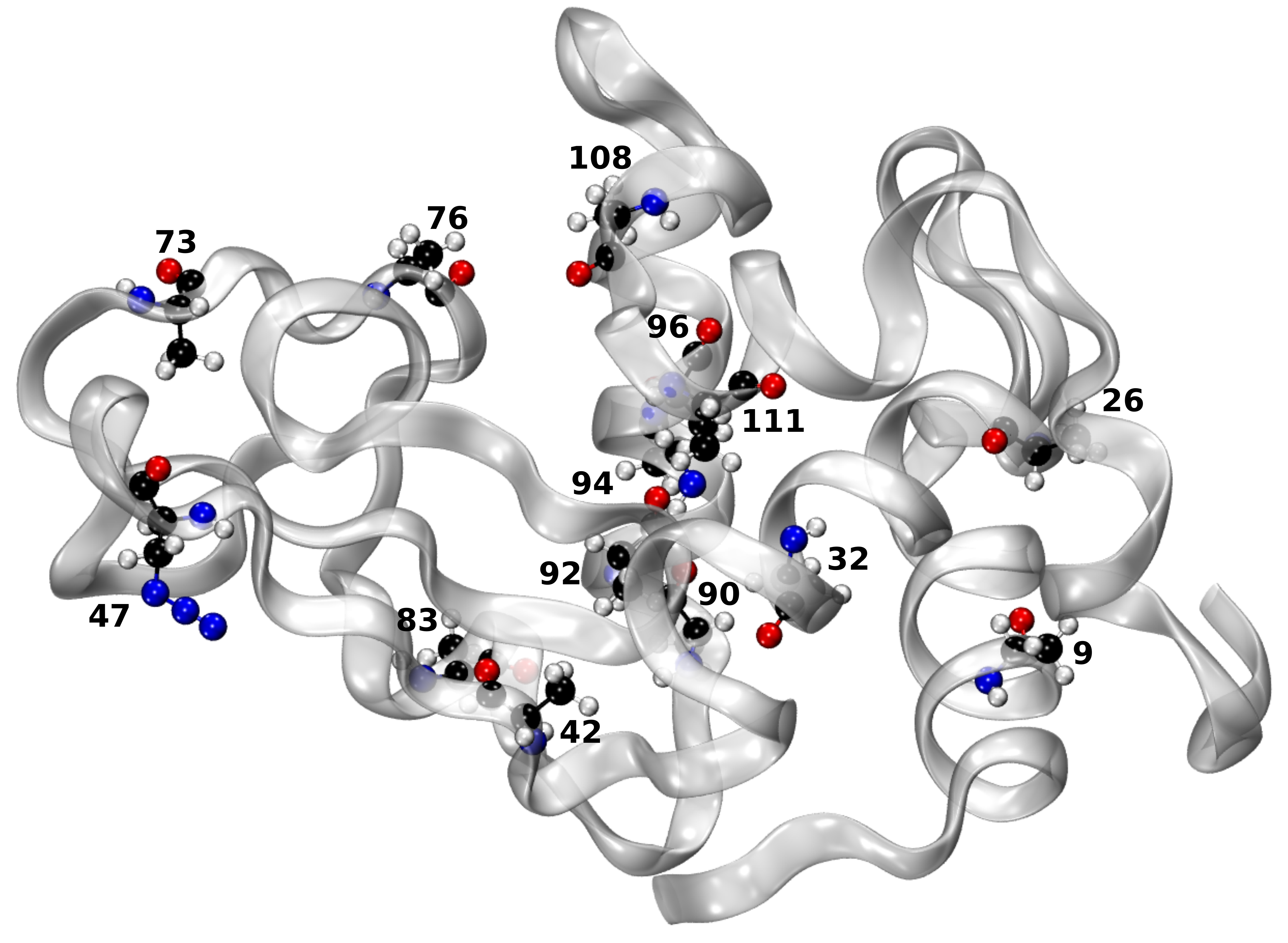}
\caption{Structure of Lysozyme with positions of Alanine residues
  indicated. The Alanine residues are at positions 9, 26, 32, 42, 47,
  73, 76, 83, 90, 92, 94, 96, 108, 111. Ala residues are displayed as
  CPK spheres and the rest of the protein structure is shown as
  NewRibbons. As an example, AlaN$_3$ is shown at residue 47.}
\label{fig:lys}
\end{center}
\end{figure}

For the Molecular Dynamics (MD) simulations of WT and modified
Lysozyme in solution, CHARMM\cite{Charmm-Brooks-2009} together with
the CHARMM\cite{charmmFF22} force field was used. A suitably modified
version of CHARMM was employed for the simulations with the
3-dimensional RKHS PESs (see below).\cite{Salehi:N3-JPCB2019} The
initial lysozyme structure was the X-ray structure
(3FE0\cite{pdblyso-2009}). Simulations of Lysozyme in TIP3P
water\cite{TIP3P-Jorgensen-1983} were carried out in a cubic box of
size $(62.1)^3$ \AA\/$^3$. Figure \ref{fig:lys} shows the structure of
the system for the present work in which N$_3^-$ is attached
individually to each of the 14 Ala residues, replacing one hydrogen
atom of the terminal CH$_3$ group. This yields Azidoalanine-modified
Lysozyme.\cite{bertozzi:2002} Compared with protein structures in
which AHA is introduced, the two modifications differ by one
CH$_2$-group.\cite{bertozzi:2002}.\\

\noindent
The systems were minimized, heated for 25 ps and equilibrated for 100
ps in the $NVT$ ensemble. Production runs, 2 ns in length, were
carried out in the $NVT$ ensemble, with coordinates saved every 5 fs
for subsequent analysis. All nonbonded interactions were treated with
a 14 \AA\/ cutoff switched at 10 \AA\/,\cite{Steinbach1994} and bonds
involving hydrogen atoms are constrained using
SHAKE\cite{SHAKE-Gunsteren-1997}.\\

\subsection{Energy Function for the Spectroscopic Probe}
For representing the 3-dimensional energy function of the N$_3^{-}$
label two strategies were pursued. First, the existing 3-dimensional
PES for N$_3^{-}$, computed at the MRCI+Q level of theory in the gas
phase, was used to describe the stretching and bending distortions of
the label attached to the CH$_2$ group of alanine.\\

\noindent
Because the N$_3^{-}$ moiety and the rest of the Ala residue are not
fully electronically decoupled, a second approach was pursued. For
this, the structure of AHA was optimized at the MP2/aug-cc-pVTZ level
of theory. Next, the structure of AHA was frozen except for the
coordinates involving the spectroscopic label. Then, a new
3-dimensional PES was computed at the pair natural orbital based
coupled cluster level
(PNO-LCCSD(T)-F12)\cite{lccsd-schwilk-2017,lccsdf12-schwilk-2017}
together with the aug-cc-pVTZ basis set\cite{DUN:JCP89} using the
MOLPRO suite of codes.\cite{molpro2} As for the gas phase
PES,\cite{Salehi:N3-JPCB2019} the \textit{ab initio} energies were
calculated in Jacobi coordinates $(R, r, \theta)$, see Figure
\ref{fig:pes}B, where $r$ is the distance between the nitrogen atoms
N1 and N2, $R$ is the distance between their center of mass and the
atom N3, and $\theta$ is the angle between $\vec{r}$ and
$\vec{R}$. The angular grid ($\theta$) used here contains 5
Gauss-Legendre quadrature points between $156^\circ$ and
$180^\circ$. The radial grids include 16 points along $r$ ranging from
0.90 to 1.51 \AA\/ and 16 points along $R$ between 1.45 and 2.12 \AA.
The PNO-LCCSD(T)-F12 level of theory was chosen as it combines
accuracy with feasibility for the present problem because recomputing
the MRCI PES for AHA is computationally intractable.\\

\noindent
For both PESs the parameters for the C-N3 stretch, the C-C-N3 and the
C-N3-N2 bend are those from Swissparam.\cite{swissparam-Zoete-2011}
All remaining parameters for the alanine residues were those of the
CHARMM force field and were not readjusted after attaching N$_3^-$ to
guarantee compatibility with the CHARMM22 force field.\\

\noindent
To carry out MD simulations for labelled Lysozyme, a continuous and
differentiable representation of the {\it ab initio} energies is
required. For this, a reproducing kernel Hilbert space-based
representation\cite{RKHS-Rabitz-1996,MM.rkhs:2017} is used. A RKHS
representation provides approximate values for a function $f(x)$ at
positions $x$, away from the grid points $x_i$. For this, the linear
problem $f(x_i) = \sum_{j} \alpha_{j} k(x_i,x_j)$ for the
1-dimensional kernels is solved which yields the coefficients
$\alpha_{j}$. There are many possible choices for the kernel functions
$k(\cdot,\cdot)$ but inverse powers of the distance have been found to
perform well for intermolecular
interactions.\cite{RKHS-Rabitz-1996,MM.heh2.1998,hutson:2000} For
multidimensional problems, tensor products of 1-dimensional kernels
can be used.\cite{hollebeek1999constructing,MM.rkhs:2017}\\

\noindent
For the present work, the 3-dimensional kernel $K$ is
\begin{equation}
K(X,X')=k^{(n,m)}(R,R')k^{(n,m)}(r,r')k^{(2)}(z,z').
\end{equation}
where $X$ stands for all dimensions involved, $r$, and $R$ are as
defined above (see also Figure \ref{fig:pes}), and
$z=\dfrac{1-\cos(\theta)}{2}$ maps the angle $\theta$ onto the
interval $[0,1]$. Reciprocal power decay kernels ($k^{(n,m)}$) with
smoothness $n=2$ and asymptotic decay $m=6$
\begin{equation}
 k^{(2,6)}(x,x') = \frac{1}{14}\frac{1}{x^7_{>}} - \frac{1}{18}\frac{x_<}{x^8_>},
\end{equation}
are used for $r$ and $R$ whereby $x_>$ and $x_<$ are the larger and
smaller values of $x$ and $x'$, respectively. For the angular degree
of freedom, a Taylor spline kernel
\begin{equation}
 k^{(2)}(z,z') = 1 + z_<z_> + 2z^2_<z_> - \frac{2}{3}z^3_<
\end{equation}
is used.\\

\noindent
Charges were calculated for the optimized structure of AlaN$_3$ at the
MP2/aug-cc-pVTZ level of theory from an NBO\cite{NBO-Foster-1980}
analysis using Gaussian\cite{g09}and scaled to maintain overall
neutrality. This yields a charge of $-0.2460e$ for the nitrogen atom
N1 attached to CH$_2$ group, $0.1607e$ for the central N2 and
$-0.0464e$ for the terminal nitrogen N3.\\

\subsection{Frequency Fluctuation Correlation Function and Lineshape}
From each production simulation, $4\times10^5$ snapshots are taken as
a time-ordered series for computing the frequency fluctuation
correlation function (FFCF) $\langle \delta \omega(0) \delta \omega(t)
\rangle$ and line shapes. Here, $\delta \omega(t) = \omega(t) -
\langle \omega(t) \rangle$ and $\langle \omega(t) \rangle$ is the
ensemble average of the transition frequency. The FFCF was determined
from instantaneous harmonic vibrational frequencies based on a normal
mode analysis.\cite{MM.insulin:2020} Normal modes were determined for
each snapshot after minimizing the structure of the N$_3$ label and
keeping the surrounding solvent frozen. Thus, frequency trajectories
$\omega_i(t)$ for label $i$ were obtained for the asymmetric stretch
vibration of N$_3$ attached to Ala. From the FFCF the line shape
function
\begin{equation}
\label{eq:gt}
g(t) = \int_{0}^{t} \int_{0}^{\tau^{'}} \langle \delta
\omega(\tau^{''}) \delta \omega(0) \rangle d\tau^{''} d\tau^{'}.
\end{equation}
is determined within the cumulant approximation. To compute $g(t)$,
the FFCF is numerically integrated using the trapezoidal rule and the
1D-IR spectra is then calculated as\cite{n3exph20-Maekawa-2004}
\begin{equation}
I(\omega) = 2 \Re \int^\infty_0
e^{i(\omega-\langle\omega\rangle)t} e^{-g(t)} e^{-\frac{t}{2T_1}}
e^{-2D_{\rm OR}t} dt,
\end{equation}
where $\langle\omega\rangle$ is the average transition frequency
obtained from the distribution, $T_1$ ($0.8 \pm 0.1$ ps) is the
vibrational relaxation time and $D_{\rm OR} = 1/6T_R$ with $T_R = 1.3
\pm 0.3$ ps is the rotational diffusion coefficient which accounts for
lifetime broadening.\cite{timedecay-Owrutsky-2003}\\

\noindent
From the FFCF, the decay time can be determined by fitting the FFCF to
a general expression\cite{hynes:2004}
\begin{equation}
  \langle \delta \omega(t) \delta \omega(0) \rangle =
  a_{1} \cos(\gamma t) e^{-t/\tau_{1}} + \sum_{i=2}^{n} a_{i}
  e^{-t/\tau_{i}} + \Delta_0
\label{eq:ffcffit}
\end{equation}
where $a_{i}$, $\tau_{i}$, $\gamma$ and $\Delta_0$ are fitting
parameters. The decay times $\tau_i$ of the frequency fluctuation
correlation function reflect the characteristic time-scales of the
solvent fluctuations to which the solute degrees of freedom are
coupled. In all cases the FFCFs were fitted to an expression
containing two decay times (i.e. $n_{\rm max} = 2$) using an automated
curve fitting tool from the SciPy library.\cite{2020SciPy-NMeth}\\

\section{Results}

\subsection{The Potential Energy Surface for the N$_3^{-}$ Label}
\begin{figure}[H]
\begin{center}
\includegraphics[width=0.8\textwidth]{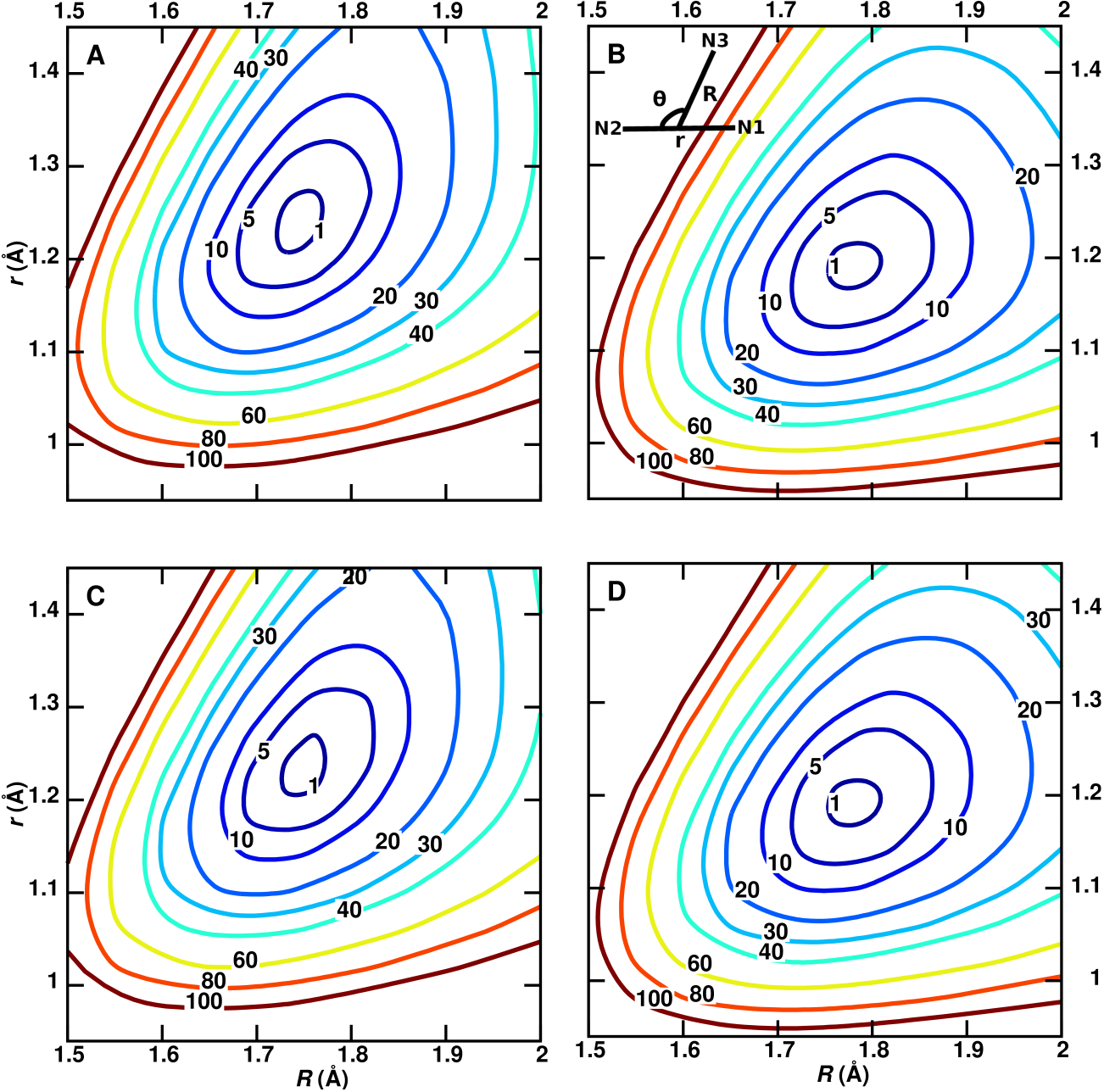}
\caption{Contour diagrams of the RKHS representations for AHA (panel
  A, PNO-LCCSD(T)-F12) and N$_3^-$ (panel B, MRCI+Q/aug-cc-pVQZ) PESs
  based on {\it ab initio} points calculated in Jacobi coordinates
  $(R, r, \theta)$ for $\theta=180^\circ$, see inset in panel
  B. Panels C and D report the corresponding CHARMM energies for
  AHA. All energies are in kcal/mol and relative to the zero of energy
  which is the minimum energy structure.}
\label{fig:pes}
\end{center}
\end{figure}

Two PESs for the energetics of the N$_3^{-}$ are considered in the
present work. One is based on earlier MRCI+Q calculations with the
aug-cc-pVTZ basis set for N$_3^{-}$ in the gas
phase\cite{Salehi:N3-JPCB2019} which was used without change for the
simulation of the AlaN$_3$ unit. The second one was the LCCSD(T) PES
for AHA which included coupling between the
N$_3^{-}$ probe and the amino acid framework. The RKHS representations
of the two PESs are reported in Figures \ref{fig:pes}A and B and the
scans within CHARMM are shown in panels C and D.\\

\noindent  
The RKHS representation of the PES for AHA was constructed from 1280
{\it ab initio} LCCSD(T)-F12 energies. An additional 230 {\it ab
  initio} energies are calculated at off-grid geometries to assess the
quality of the RKHS representation. Figure S1 shows
the correlation between the reference energies and the RKHS with a
correlation coefficient of $R^{2} = 0.9999$ and the root mean squared
error is 0.38 kcal/mol. This confirms the high quality of the RKHS
PES.\\

\noindent
Figure \ref{fig:pes}A and B report the RKHS interpolation of the {\it
  ab initio} calculated energies whereas Figures \ref{fig:pes}C and D
are from scanning the $r$ and $R$ coordinates for AHA in the gas phase
in CHARMM. Comparing the PNO-LCCSD(T)-F12 PES (Figure \ref{fig:pes}A)
with that at the MRCI+Q level of theory (Figure \ref{fig:pes}B) shows
that the minima for the two are slightly displaced ($r = 1.19$ \AA\/
vs. $r = 1.24$ \AA\/ and $R = 1.77$ \AA\/ vs. $R = 1.76$
\AA\/). Furthermore, the LCCSD(T) PES is steeper along both, the $r$
and $R$ coordinates, which pushes the respective vibrations up
compared with the MRCI+Q PES, see Figure S2. Differences
between the two PESs are due to both, the methods (MRCI+Q
vs. PNO-LCCSD(T)-F12) and the model system (N$_3^-$ vs. AHA)
considered. Comparing the isolated, gas-phase PESs (panels A and B)
with those for AlaN$_3$ (panels C and D) indicates that the PESs are
close but not identical due to coupling between the spectroscopic
probe and the alanine residue.\\

\noindent
In the following, all MD simulations were carried out with the
PNO-LCCSD(T)-F12 PES as it yields harmonic frequencies for AlaN$_3$
around 2110 cm$^{-1}$ cm$^{-1}$ (see Table \ref{tab:ffcffit}) which is
consistent with those experimentally observed for the replacement of
AHA\cite{stock:2018} in PDZ2 domain at 2114 cm$^{-1}$ and for
AlaN3\cite{alan3-jcp-2015} in H$_2$O at 2116 cm$^{-1}$,
respectively. Moreover, the influence of the covalent bonding to the
Alanine residue is included in the construction of the potential
energy surface. Additional refinements of the PES would, in principle,
be possible through morphing\cite{JMB91morphing,MM.morph:1999} but
were not deemed necessary for the present work which is mainly
concerned with the differential dynamics, i.e. the relative positional
sensitivity, and spectroscopy for the same label at different
positions along the polypeptide chain.\\

\subsection{Structural Dynamics}
\begin{figure}[H]
\centering
\includegraphics[width=0.4\textwidth,angle=-90]{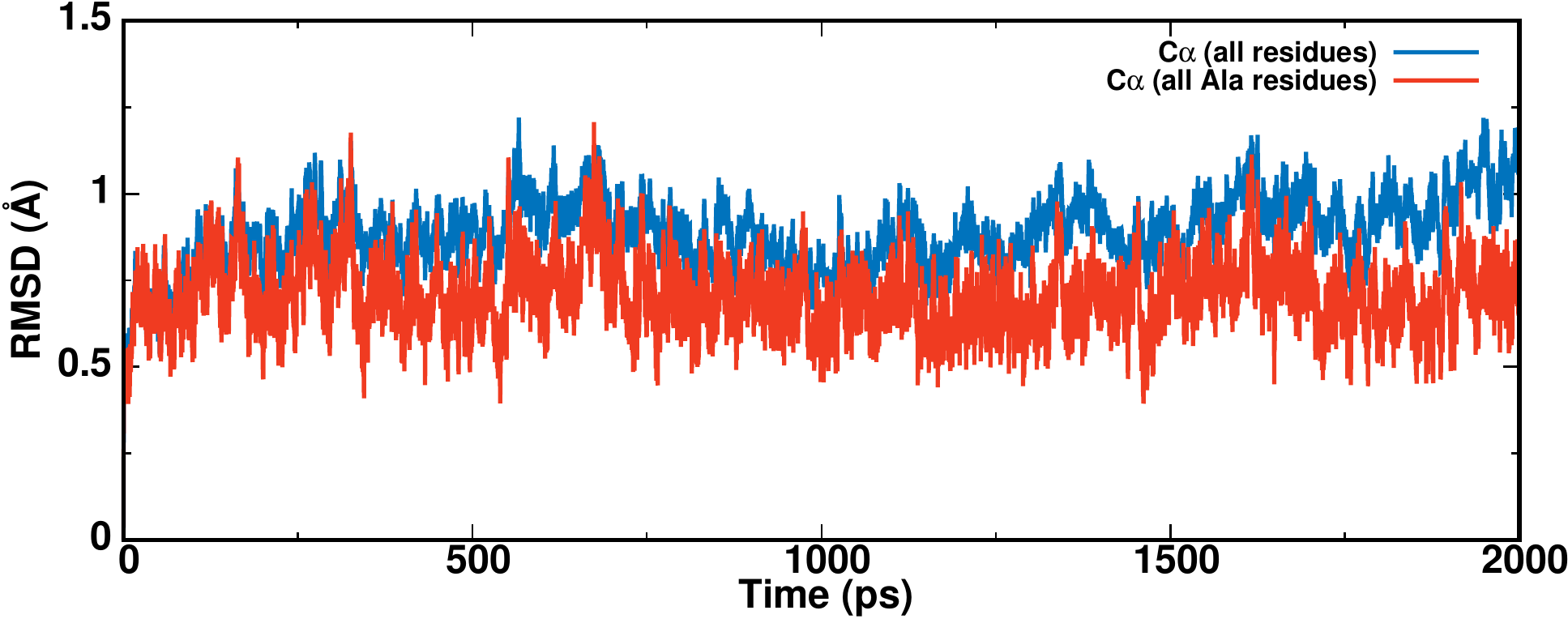}
\caption{The structural RMSD for the C$_{\alpha}$ atoms from all
  residues (blue) and for the 14 Ala residues (red) specifically for
  Ala76N$_3$.}
\label{fig:rmsd_lys}
\end{figure}

For the structural dynamics first the root mean squared deviation
(RMSD) of unmodified and modified Lysozyme in solution compared with
the starting X-ray structure as the reference is analyzed. For this,
the RMSD of all C$_{\alpha}$ atoms was considered. Figure
\ref{fig:rmsd_lys} shows the RMSD for all C$_{\alpha}$ atoms (blue)
and those for the 14 Alanine residues (red) specifically from the 2 ns
simulation of the modified protein at position Ala47. For the WT
protein similar RMSD values are reported in Figure S3. The RMSD values fluctuate below or around 1
\AA\/ which is indicative of a stable simulation. This suggests that
attaching a N$_3^-$ label to Ala has an insignificant effect on the
structural dynamics of Lysozyme, consistent with earlier findings for
the PDZ domain for which also a minimally invasive effect was
reported.\cite{stock:2018}\\

\subsection{Vibrational Spectra and Frequency Correlation Functions}
\begin{figure}[H]
\begin{center}
\includegraphics[width=0.38\textwidth,angle=-90]{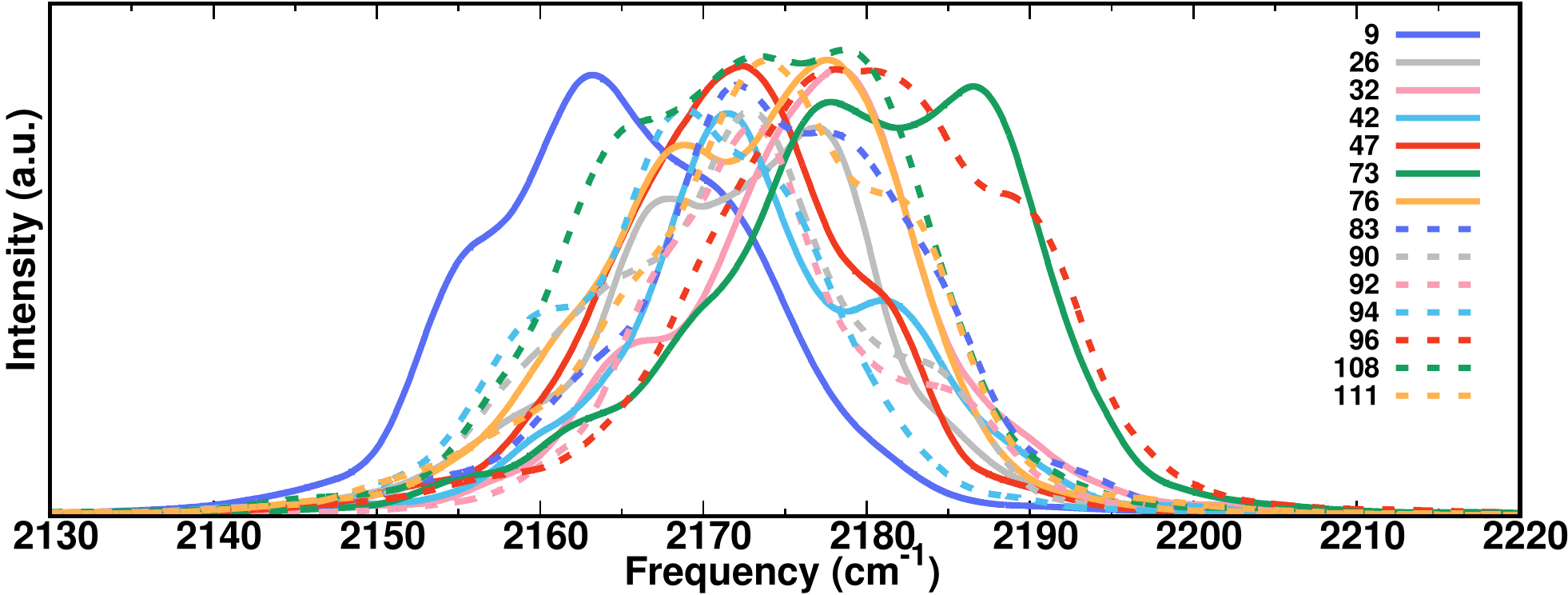}
\caption{Power spectrum based on the N$2$--N$3$ separation for all
  modified AlaN$_3$ residues. The position of the frequency maxima
  differ for most of the AlaN$_3$ labels and cover a range between
  2160 and 2180 cm$^{-1}$.}
\label{fig:avgps}
\end{center}
\end{figure}

First, the power spectra and frequency trajectories for the asymmetric
stretch of the azide label attached to all 14 alanine residues are
presented. The power spectra as determined from the Fourier transform
of the N2-N3 distance correlation function are shown in Figure
\ref{fig:avgps}A for all AlaN$_3$ from 2 ns production runs. The peak
maxima $\omega_{\rm max}$ cover a range of $\sim 20$ cm$^{-1}$
(between 2160 and 2180 cm$^{-1}$) and the full widths at half maximum
(fwhm) of the spectra are around 20 cm$^{-1}$. Hence, although the
same energy function was used for all modified AlaN$_3$ moieties,
their power spectra differ depending on the position of the modified
Ala residue along the polypeptide chain.\\

\noindent
The power spectra reported in Figure \ref{fig:avgps} are also
representative of the infrared spectrum as shown in Figure
S4. The top panel of Figure S4
reports the power spectrum and peak positions of all three modes for
Ala47N$_3$ with the asymmetric stretch centered around 2170 cm$^{-1}$,
the symmetric stretch at 1333 cm$^{-1}$ and the bending mode at 610
cm$^{-1}$. The bottom panel of Figure S4
demonstrates that the infrared spectrum (IR) determined from the
dipole autocorrelation function supports the peak positions found from
the power spectrum to within 2 cm$^{-1}$.\\

\noindent
Next, the frequency trajectories $\omega_i(t)$ for each of the
spectroscopic probes $i$ from $4 \times 10^5$ snapshots were
determined from instantaneous normal mode calculations. From the
frequency time series the frequency fluctuation correlation functions
(FFCFs) are obtained. They contain valuable information about the
environmental dynamics around each site $i$, i.e. the azide probes of
the various Ala residues considered.\\

\begin{figure}[H]
\centering
\includegraphics[width=0.9\textwidth]{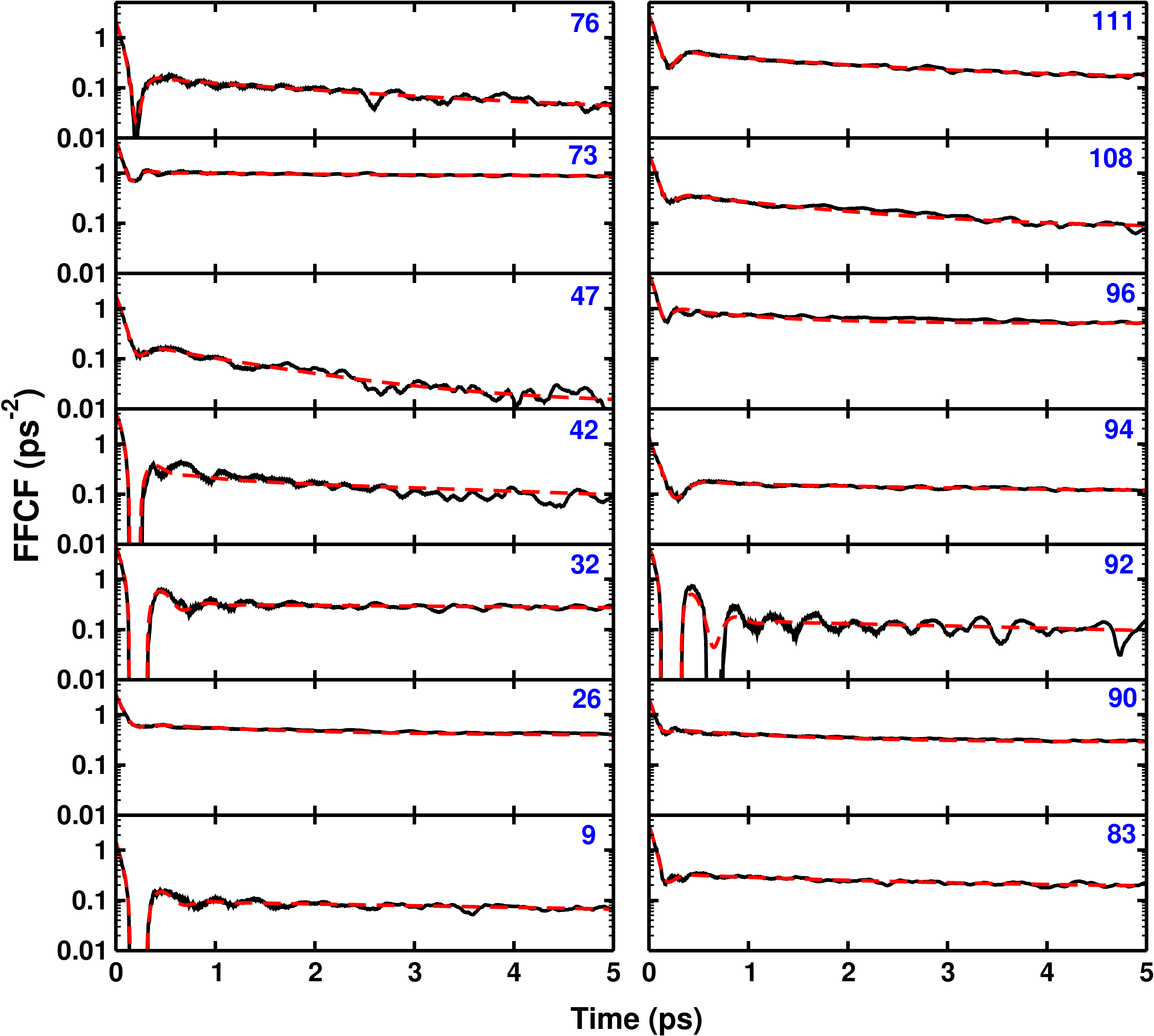}
\caption{FFCFs from correlating the instantaneous harmonic frequencies
  for all 14 AlaN$_3$ in Lysozyme. The labels in each panel refer to
  the alanine residue which carries the azide label. Black traces are
  the raw data and red dashed lines the fits to
  Eq. \ref{eq:ffcffit}. The $y-$axis is logarithmic.}
\label{fig:ffcffit}
\end{figure}

\noindent
The FFCFs, shown in Figure \ref{fig:ffcffit}, are fitted to
Eq. \ref{eq:ffcffit} with a parametrization motivated by the overall
shape of the FFCF.\cite{hynes:2004} This functional form has also been
used in previous work.\cite{hynes:2004,skinner:2006,MM.cn:2013} It is
an extension of the typical multiexponential decay, which is
traditionally employed\cite{hamm:1998} to capture an anticorrelation
at short times ($t < 1$ ps). Figure \ref{fig:ffcffit} provides a
comparison between the raw data (black) and the fits (red) and Table
\ref{tab:ffcffit} reports the corresponding fitting parameters.\\

\noindent
The shape of the FFCFs can differ appreciably. Some of them display a
pronounced minimum at short correlation times ($t \sim 0.1$ ps)
whereas others do not. This feature has also been found in previous
simulations\cite{skinner:2006} and has been related to the strength of
the interaction between the infrared probe and its
environment.\cite{hynes:2004,MM.cn:2013,NMA2dir-MM-2014} Several of
the FFCFs show one (Ala9, Ala32, Ala42, Ala76, Ala94, Ala96, Ala108,
Ala111) or even two (Ala92) recurrences at short correlation
times. For the remaining Alanine residues this feature is less
pronounced (Ala47, Ala73, Ala83, Ala90) or entirely absent
(Ala26). Similarly, some of the FFCFs exhibit clear static components
$\Delta_0 \simeq 0.5$ ps$^{-2}$ (Ala26, Ala73, Ala96) whereas the
remaining ones decay to zero on the $\sim 10$ ps time scale. With
respect to the correlation times, the fast correlation is generally
$\tau_1 \sim 0.1$ ps whereas the long time scale ranges from $\tau_2 =
1.1$ ps to $\tau_2 < 13$ ps, see Table \ref{tab:ffcffit}. Typically,
the amplitude of the fast decay is one order of magnitude larger than
that of the slow contribution (Table \ref{tab:ffcffit}). Hence, the
characteristics of the FFCFs vary considerably depending on the
position at which the Alanine residue is located along the
polypeptide chain. This suggests that AlaN$_3$ is a positionally
sensitive probe to provide quantitative information about the local
dynamics of a protein.\\

\begin{table}[H]
\footnotesize
\centering
\caption{Parameters obtained from fitting the FFCF to
  Eq. \ref{eq:ffcffit} for INM frequencies for all different AlaN$_3$
  residues in lysozyme. Average frequency $\langle\omega\rangle$ of
  the asymmetric stretch in cm$^{-1}$, the amplitudes $a_1$ to $a_3$
  in ps$^{-2}$, the decay times $\tau_1$ to $\tau_3$ in ps, the
  parameter $\gamma$ in ps$^{-1}$, the offset $\Delta_0$ in ps$^{-2}$,
  and the conformationally averaged local hydrophobicity (LH).}
\centering
\begin{tabular}{r|c|crc|cc|c||c}
\hline\hline
Res& $\langle\omega\rangle$ & $a_{1}$ & $\gamma$ &$\tau_{1}$ &$a_{2}$ &$\tau_{2}$ & $\Delta_0$& LH\\
\hline
\textbf{9} &2103.7& 1.17 & 13.16 &  0.129 & 0.07 &  7.62 &   0.02 & 0.33\\
\hline
\textbf{26} &2107.9& 1.82 & 8.69 & 0.079 &   0.24 &   1.73 &   0.41 &1.91 \\
\hline
\textbf{32} &2112.8&  3.60 & 13.39 &  0.164 &   0.13 &  13.05  &   0.19&0.11 \\
\hline
\textbf{42} &2111.4&  3.57 & 14.08 & 0.104 &   0.32 &  2.15 &   0.04&0.95\\
\hline
\textbf{47}&2107.5&  1.54 & 9.99 &  0.081 &  0.21 &   1.17 &  0.01&1.21 \\
\hline
\textbf{73}&2114.6&  3.18 & 15.07 &   0.080 &   0.24 &   3.61 &    0.81&1.34 \\
\hline
\textbf{76} & 2107.1&  1.94 & 11.94&   0.084 &   0.15 &   2.77 &    0.02&0.40\\
\hline
\textbf{83} &2109.2&  2.75 & 11.48 &  0.069 &  0.17 &   2.11 &     0.18 &1.59\\
\hline
\textbf{90} &2107.5&  1.57 & 11.71 &   0.056 &  0.22 &   1.42 &    0.30 &1.13\\
\hline
\textbf{92} &2110.4&  4.08 & 14.02 &   0.184 &   0.10 &  1.33 & 0.10 &0.58\\
\hline
\textbf{94} &2104.8&  1.05 & 9.36  &  0.100 &   0.08 &   5.89 &   0.09 &1.17\\
\hline
\textbf{96}& 2116.3&  3.74 & 13.33 &  0.077&   0.43 &   2.19 &   0.48 &0.44\\
\hline
\textbf{108} &2108.4& 1.83 & 10.77 &  0.084 &   0.35 &   2.43 &    0.04 &0.54\\
\hline
\textbf{111} &2110.6& 2.22 & 13.13 &  0.090 &   0.41 &    2.27 &   0.12&1.50\\
\hline\hline
\end{tabular}
\label{tab:ffcffit}
\end{table}

\begin{figure}[H]
\begin{center}
\includegraphics[width=0.38\textwidth,angle=-90]{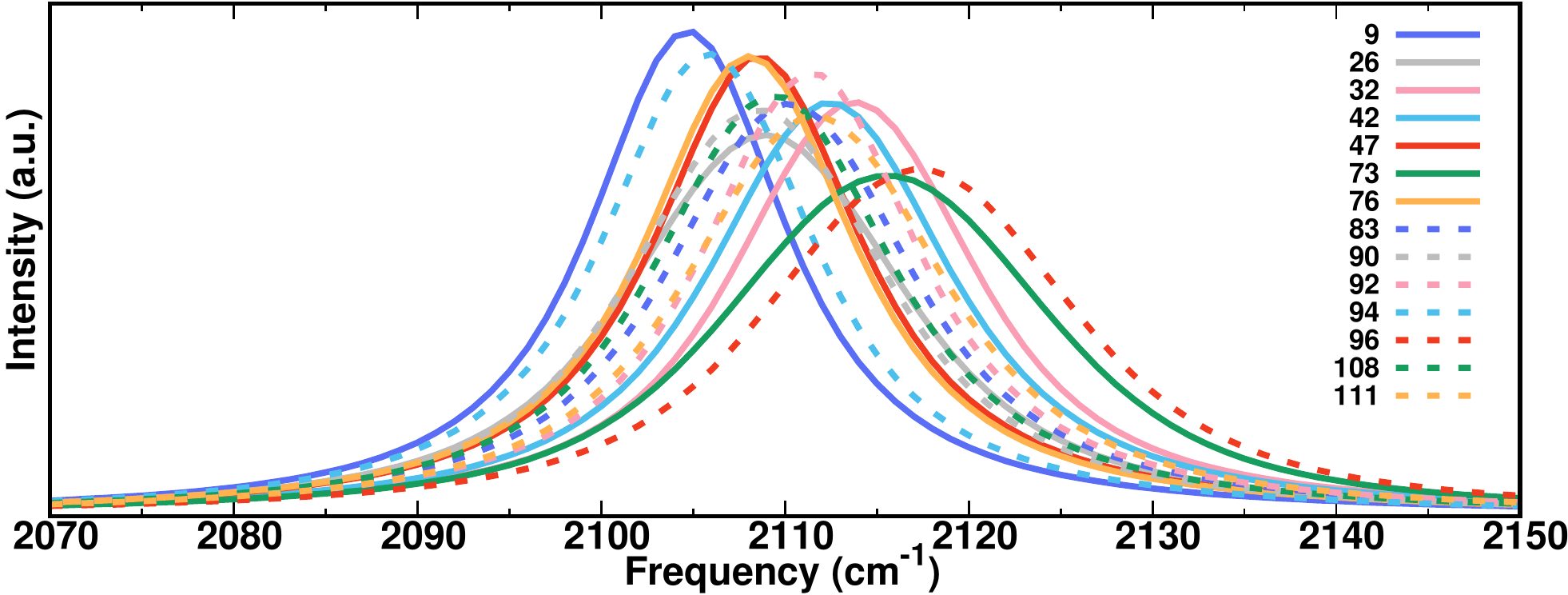}
\caption{1D IR spectra for all 14 AlaN$_3$ residues in Lysozyme. For
  the IR lineshape the raw FFCF from the INM analysis was numerically
  integrated to give $g(t)$ from which the 1D lineshape is obtained.}
\label{fig:ir}
\end{center}
\end{figure}

\noindent
Numerical integration of $g(t)$ and using Eq. \ref{eq:ffcffit} yields
the 1-dimensional IR spectra for each label based on instantaneous
normal modes, see Figure \ref{fig:ir}. Similar to the power spectra,
the center frequencies cover a range of $\sim 15$ cm$^{-1}$, with
center frequencies of 2104 cm$^{-1}$ for Ala9N$_3$ and 2116 cm$^{-1}$
for Ala96N$_3$, and the fwhm ranges from 13 to 21 cm$^{-1}$. Also, the
$\omega_{\rm max}$ for Ala9N$_3$ (blue solid line in Figures
\ref{fig:avgps} and \ref{fig:ir}) is lowest in frequency and those for
Ala96N$_3$ (dashed red) and Ala73N$_3$ (solid green) are highest from
the power spectra and the INM lineshapes, respectively. The blue shift
of the power spectra compared with those from INM for the symmetric
and asymmetric stretch modes was already found for N$_3^-$ in
solution.\cite{Salehi:N3-JPCB2019} The magnitude of this shift is
larger in the present case probably due to coupling between the
spectroscopy probe and the amino acid it is attached to.\\

\noindent
An alternative to instantaneous normal modes is to obtain
instantaneous frequencies from solving the 1- or 3-dimensional nuclear
Schr\"odinger equation. For this, the corresponding 1- or 3-d PES is
scanned for a given snapshot with frozen
environment\cite{MM.cn:2013,MM.insulin:2020,MM.rev.jcp:2020} and
represented as a RKHS. This is a computationally much more demanding
approach, in particular in 3 spatial
dimensions.\cite{Salehi:N3-JPCB2019} Here, the 1-dimensional PES along
the asymmetric stretch motion was mapped out for $4 \times 10^5$
snapshots and the nuclear Schr\"odinger equation was solved. Then, the
FFCF was again determined and fit to Eq. \ref{eq:ffcffit}, see Figure S5. From this, the
1-dimensional IR lineshape was determined, see solid lines in Figure
S6. This was done for Ala90N$_3$ and Ala94N$_3$. As was
found for the 1-d lineshapes from INM, the frequency maximum for
Ala90N$_3$ is shifted to the blue relative to Ala94N$_3$ but the shift
is smaller (1 cm$^{-1}$ vs. 3 cm$^{-1}$).\\

\noindent
Recently, the ``INM'', ``scan'' and ``map'' approaches have been
compared for insulin monomer and dimer.\cite{MM.insulin:2020} It was
found that the ``INM'' and ``scan'' approaches yield comparable 1-d
infrared spectra for the amide-I bands and conclusions drawn from the
spectra concerning monomeric and dimeric insulin are
consistent. Nevertheless, the two approaches can differ in the
absolute frequencies as is also found in the present case.\\

\noindent
Typically, spectroscopic work has used AHA as an infrared label
instead of azidoalanine as used here. To quantify the difference
between AlaN$_3$ and AHA, residue Ala47 has also been replaced by AHA
through inserting an additional CH$_2$ group before the N$_3^{-}$
label. The parametrization of the CH$_2$ group is identical to that
already used for alanine. Then, a 2 ns simulation for AHA in water was
carried out and the IR spectrum was determined from an INM analysis,
see Figure S7. It is found that the position of the
frequency maximum for the asymmetric stretch of the azide label
differs by less than 1 cm$^{-1}$ from that with AlaN$_3$ which
confirms that for IR spectroscopy, the two systems are very similar.\\

\section{Solvent Structure and Dynamics}
\begin{figure}[H]
  \centering
    \includegraphics[width=\textwidth]{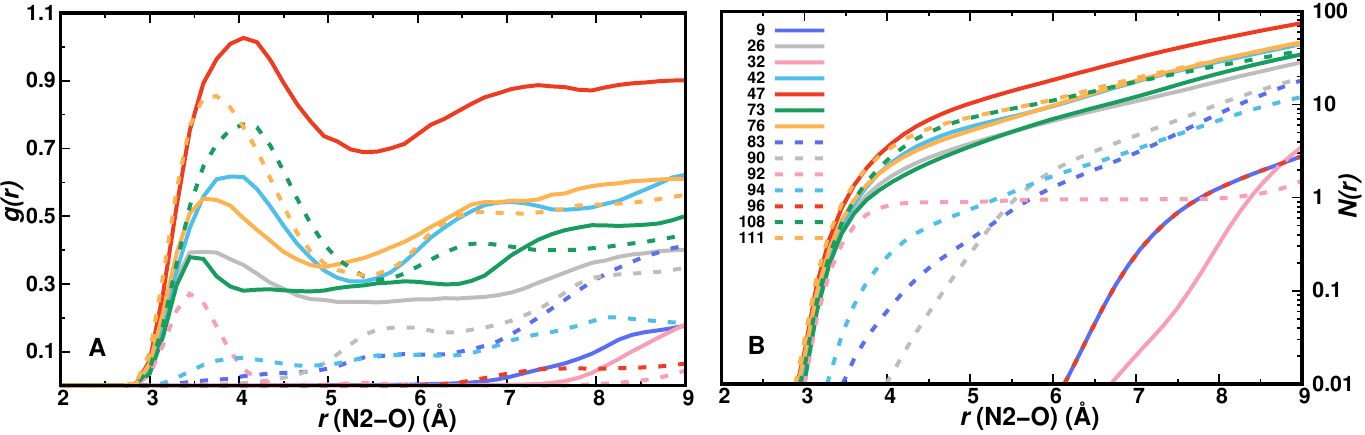}
    \caption{The radial distribution function $g(r)$ (panel A) and the
      number of water oxygen atoms $N(r)$ (panel B) between O of water
      and N2 of AlaN$_3$ for all alanine residues from the 2 ns
      production simulations. The color code for the lines is given in
      panel B.}
    \label{fig:grnr}
\end{figure}

Next, the solvent structuring around the modification sites is
characterized. This also provides the information for an attempt to
relate the spectral signatures (position of the frequency maximum,
characteristics of the FFCFs) for the azide labels at different
positions along the polypeptide chain with structural features and
environmental properties. For this, the solvent structure around each
of the 14 AlaN$_3$ probes was analyzed. First, the radial distribution
functions $g(r)$ were computed along all production simulations for
the 14 modified proteins, see Figure \ref{fig:grnr}. The distance
analyzed was the separation between the water-oxygen atom (O$_{\rm
  W}$) and the middle nitrogen (N2) of the N$_3$ probe in
AlaN$_3$. The corresponding running coordination number $N(r)$ is
\begin{equation*}
N(r)= 4\pi \int^r_0   r^2 g(r) \rho dr 
\end{equation*}
where $\rho$ is the pure water density (Figure \ref{fig:grnr}B). As is
shown in Figure \ref{fig:grnr}, the $g(r)$ and $N(r)$ differ for the
14 modification sites.\\

\noindent
For some of the residues (Ala26, Ala42, Ala47, Ala73, Ala76, Ala108,
Ala111; Set1) the $g(r)$ exhibits a pronounced first maximum at $3.5
\leq r_{\rm max1} \leq 4$ \AA\/ whereas for the remaining labels
(Ala9, Ala32, Ala83, Ala90, Ala92, Ala94, Ala96; Set2) such a first
maximum is largely absent. This suggests that the residues in Set1 are
solvent exposed whereas those in Set2 are not. The total number $N(r)$
of water molecules within a distance $r$ supports this, see Figure
\ref{fig:grnr}B. Up to a distance of 5 \AA\/, which is typically the
extent of the first hydration shell, residues in Set1 contain 10 or
more water molecules whereas those belonging to Set2 have not more
than 1 water molecule in their vicinity.\\

\begin{figure}[H]
  \centering
    \includegraphics[width=\textwidth]{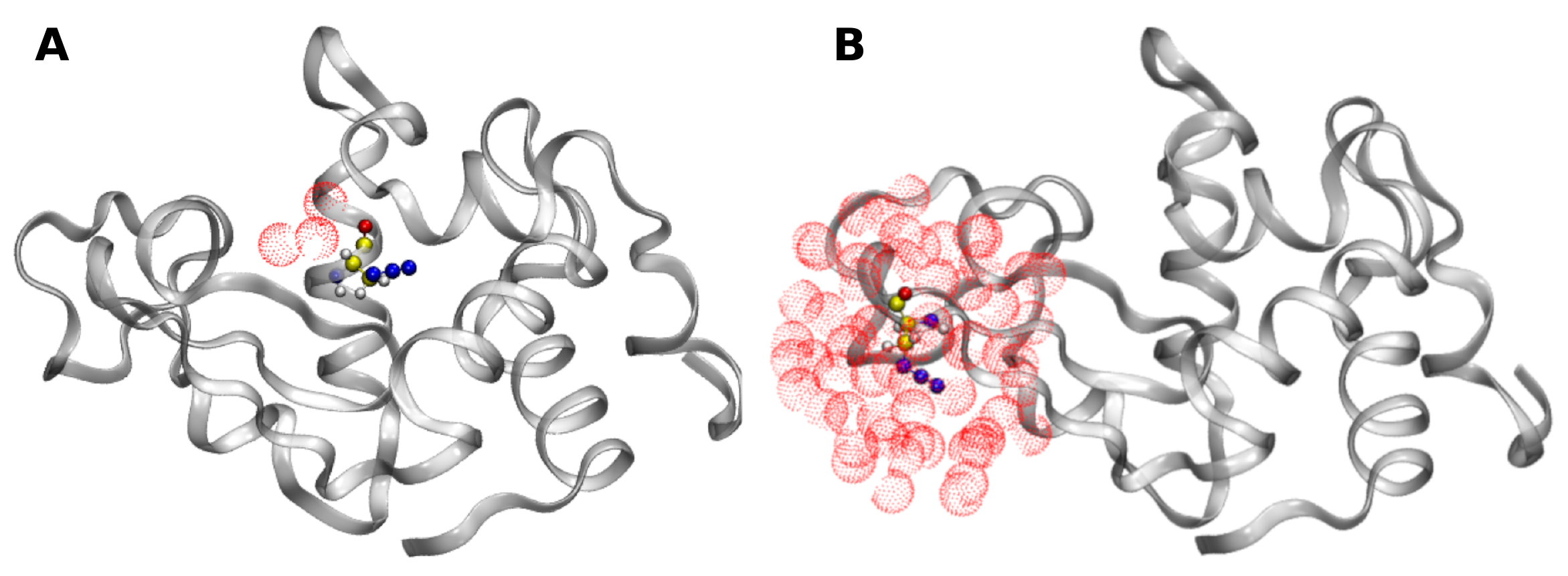}
    \caption{Solvent distribution based on the water-oxygen atoms
      within 7 \AA\/ of any atom of residues Ala96N$_3$ (panel A) and
      Ala47N$_3$ (panel B). The small and large hydration spheres are
      consistent with the $g(r)$ and $N(r)$ reported in Figure
      \ref{fig:grnr}.}
    \label{fig:solv_dist}
\end{figure}

\noindent
A structural illustration for this observation is given in Figure
\ref{fig:solv_dist} which reports all water molecules within 7 \AA\/
of Ala47N$_3$ (belonging to Set1) and Ala96N$_3$ (belonging to
Set2). Consistent with Figure \ref{fig:grnr} only 3 water molecules
are within the cutoff radius of atom N2 of Ala96N$_3$ whereas the
hydration shell of Ala47N$_3$ is extensive.\\

\noindent
Another measure to quantify the solvent exposure of amino acids is to
determine the time dependent quantity,
$\delta\lambda_{\textrm{phob}}^{(r)}(t)$, which is referred to as the
local hydrophobicity (LH) of residue $r$ at time
$t$.\cite{Willard-jctc-2018,Willard-jpcb-2018} This measure is based
on analyzing the occupation and orientational statistics of surface
water molecules at the protein/water interface, given by the three
dimensional vector $\vec{\kappa} = (a, \cos \theta_{\rm OH1}, \cos
\theta_{\rm OH2})$. Here, $a$ is the distance of the water oxygen atom
to the nearest atom of residue $r$\cite{MM.hb:2020}, and $\theta_{\rm
  OH1}$ and $\theta_{\rm OH2}$ are the angles between the water OH1
and OH2 bonds and the interface normal. More specifically, the local
hydrophobicity (LH) is $\delta \lambda_\mathrm{phob}^{(r)}(t) =
\lambda_\mathrm{phob}^{(r)}(t) - \langle \lambda_\mathrm{phob}
\rangle_0$, where
\begin{equation}
    \lambda_\textrm{phob}^{(r)}(t) = -\frac{1}{{\sum_{a=1}^{N_a(r)}
        N_w(t;a)}}\sum_{a=1}^{N_a(r)} \sum_{i=1}^{N_w(t;a)} \ln{\left[
        \frac{P(\vec{\kappa}^{(i)}(t)|\textrm{phob})}{P(\vec{\kappa}^{(i)}(t)|\textrm{bulk})}
        \right]}
\label{eq:lh}
\end{equation}
and $\langle \lambda_\mathrm{phob} \rangle_0$ is the ensemble average
sampled from the ideal hydrophobic reference system (see below). The
summation over $N_a(r)$ involves all atoms in residue $r$ and the
summation over $N_w(t;a)$ includes all water molecules within a
cut-off of 6\AA\/ of atom $a$ at time $t$.\cite{MM.hb:2020} The vector
$\vec{\kappa}^{(i)}(t)$ describes the orientation (see above) of the
$i$th water molecule in the sampled population.\\

\noindent
The distribution $P(\vec{\kappa}^{(i)}(t)|\textrm{phob})$ is
determined for a reference hydrophobic reference system ('phob'),
whereas $P(\vec{\kappa}^{(i)}(t)|\textrm{bulk})$ is determined from
the actual simulations ('bulk').\cite{Willard-jctc-2018} As the
quantity LH includes both, the distance $a$ of the water molecules
from the interface and the orientation of a specific water molecule
($\theta_{\rm OH1}, \theta_{\rm OH2}$), LH can be considered as a
generalization of the radial distribution function $g(r)$. The local
hydrophobicity is a measure of the statistical similarity of the
sampled configurations to that of an ideal hydrophobic reference
system. When sampled configurations
$P(\vec{\kappa}^{(i)}(t)|\textrm{bulk})$ are dissimilar to the
hydrophobic reference system, this indicates that the site $r$
considered is less hydrophobic, i.e. rather hydrophilic and vice
versa. In other words, $\delta \lambda_\mathrm{phob}^{(r)}(t) \approx
0$ for a hydrophobic environment around residue r, whereas $\delta
\lambda_\mathrm{phob}^{(r)}(t)$ significantly larger than zero, the
environment is
hydrophilic.\cite{Willard-jctc-2018,Willard-jpcb-2018,MM.hb:2020} In
previous work\cite{MM.hb:2020}, sustained values of $\delta
\lambda_\mathrm{phob}^{(r)} > 0.5$ were considered indicative of
hydrophilicity. The magnitude of such a cutoff may, however, be
somewhat system-dependent.\\

\begin{figure}[H]
\centering
\includegraphics[width=0.4\textwidth,angle=-90]{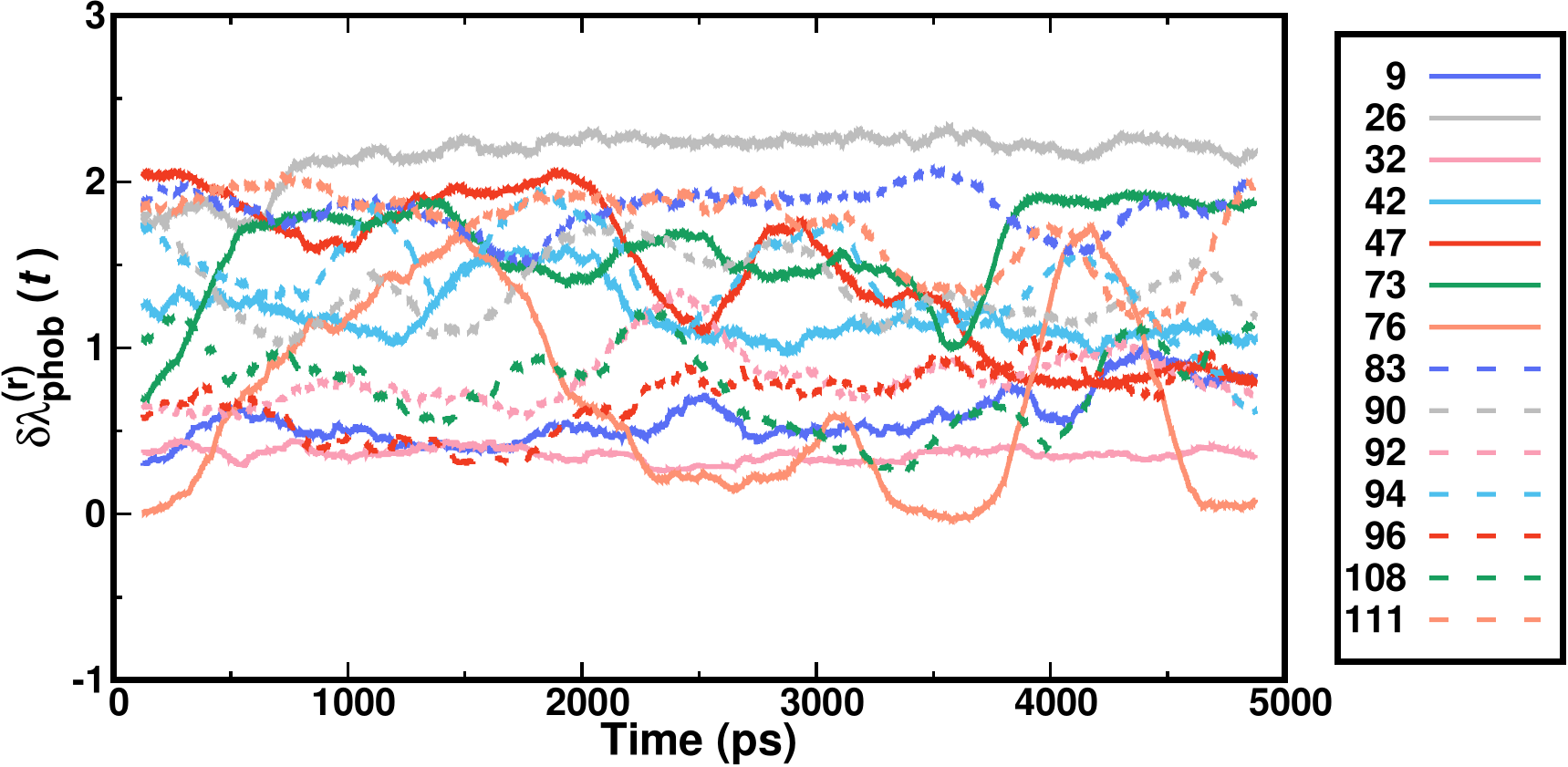}
\caption{Local hydrophobicity as a function of time for all alanine
  residues from the simulation of WT Lysozyme. The LH coefficient was
  determined from Eq. \ref{eq:lh}. Values of $\delta \lambda_{\rm
    phob} \approx 0$ indicate a hydrophobic environment of the site
  considered\cite{Willard-jpcb-2018,MM.hb:2020} whereas values around
  2 point towards a hydrophilic site.}
\label{fig:hyd5}
\end{figure}

\noindent
Figure S8 gives an overview of the average LH per residue
and the fluctuations around the average for WT Lysozyme. The Alanine
residues (in red) are found to include both, low and high values for
LH, representative of more hydrophobic and hydrophilic environments,
respectively. The change in LH as a function of simulation time (over
2 ns) for WT (blue) and N$_3^{-}$ labelled (red) Lysozyme for Ala76 is
reported in Figure S9. Without spectroscopic label the
Ala-residue is rather hydrophilic on average whereas with the label
attached it is more hydrophobic (less hydrophilic). On the other hand,
the LH can have a rather pronounced time-dependence, see Figure
\ref{fig:hyd5} (solid orange line for Ala76) from the 5 ns simulation
of WT Lysozyme. Thus, attaching the N$_3^{-}$ label to Ala may
modulate recruitment or displacement of solvent molecules.\\

\section{Discussion and Conclusion}
The present findings confirm that azide attached to alanine residues
in Lysozyme is a structurally minimally invasive, specific infrared
label to quantitatively probe the local dynamics around the
modification site. This has already been reported for the PDZ2
domain.\cite{stock:2018} Similar to the situation in insulin monomer
and dimer, for which the amide-I vibration was
found\cite{MM.insulin:2019,MM.insulin:2020} to cover a range of $\sim
20$ cm$^{-1}$, attaching azide to give AlaN$_3$ spans a comparable
frequency range but in a region of the infrared spectrum (around 2100
cm$^{-1}$) that is typically ``empty''. Together with their minimal
impact on the overall protein structure (see Figure
\ref{fig:rmsd_lys}), and the still favourable extinction
coefficient\cite{hamm:2012}, such modifications bear great potential
to resolve the structural dynamics of proteins and protein-ligand
complexes at a molecular level. Studies that provide structural and
spectroscopic information at the same time are of great interest for
characterizing potential ligand-binding sites and for functional
studies of protein allostery.\\

\begin{figure}[H]
\begin{center}
  \includegraphics[width=\textwidth]{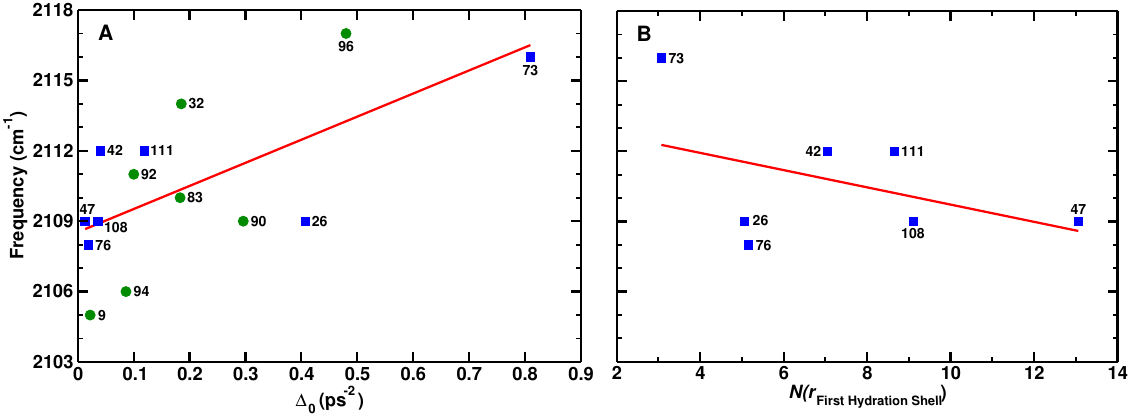}
  \caption{Correlation between the maximum of the 1-d lineshape from
    INM and he static offset $\Delta_0$ of the FFCF (panel A) and the
    maximum of the 1-d lineshape from INM and the number of water
    molecules in the first hydration shell (panel B) for the residues
    that has been considered to be rather ``water exposed''
    (Set1). Residues of Set1 are shown as blue squares and those of
    Set2 as green circles. The solid line is an empirical linear fit
    and suggests that, typically, for more blue shifted frequency
    maxima the static component increases while the number of water
    molecules in the first hydration shell decreases, i.e.  with
    increasing hydration, $\omega_{\rm max}$ shifts typically to the
    red for alanine residues in Set1.}
\label{fig:correl.ir}
\end{center}
\end{figure}

\noindent
It is of interest to delineate whether correlations can be found
between structural and spectroscopic characteristics analyzed in the
present work. As the dynamics is coupled and involves a potentially
complicated superposition of different structural substates, no
``simple'' or ``obvious'' correlations are expected. Rather and at
best, discovering trends can be expected from such an analysis. One
example is shown in Figure \ref{fig:correl.ir}B which reports the
relationship between the number of water molecules in the first
hydration shell (see also Figure \ref{fig:grnr}) and the position of
the frequency maximum $\omega_{\rm max}$ from the 1-d lineshape
determined from the instantaneous normal mode analysis. Typically,
with increasing hydration, the position of $\omega_{\rm max}$ shifts
to the red. Similarly, the magnitude of the static offset $\Delta_0$
of the FFCF is related to $\omega_{\rm max}$ in that larger values of
$\Delta_0$ are associated with a blue shift of the position of the
frequency maximum, see Figure \ref{fig:correl.ir}A.\\

\noindent
Spectroscopic probes to characterize the local environment of a
protein provide valuable information about local hydration. This is of
particular relevance given the findings that individual water
molecules can play decisive roles in protein function. For example, in
HIV-I protease\cite{erickson:1995,prashar:2009} a single catalytic
water molecule was located in the active site of the protein or for
insulin\cite{MM.insulin:2019,MM.insulin:2018} individual water
molecules were found to attack the dimerization interface to reduce
the thermodynamic stability of the dimer by a factor of
two. Similarly, water molecules have been reported to play essential
roles in protein folding,\cite{weik:2015} and for
function.\cite{pocker:2000,zewail:2004} Thus, probing and
characterizing the local solvent environment of particular regions of
a protein can provide important insights into functional aspects of
proteins.\\

\noindent
The utility of infrared spectroscopy to study the strength of
protein-ligand complexes has been proposed\cite{Suydam.halr.sci.2006}
and explicitly demonstrated from molecular dynamics simulations for
cyano-substituted benzene in lysozyme.\cite{MM.lys:2017} Using AHA as
a probe, it was reported that unbound and ligand-bound PDZ2 differ in
that the frequency correlation function for the two systems decay to
different levels at longer correlation times. Similarly, infrared
spectroscopy is also a sensitive probe - both, in terms of
spectroscopy and dynamics - to characterize protein-protein
interactions.\cite{MM.insulin:2020} Together with experimental
studies,\cite{tokmakoff:2016.2,tokmakoff:2020,dinner:2020} such
efforts pave the way for functional, in vivo studies of protein-ligand
and protein-protein association.\cite{smith:2013}\\

\noindent
In conclusion, the present work provides a comprehensive analysis of
the spectroscopy and dynamics of azide-labelled alanine in
Lysozyme. The results demonstrate that AlaN$_3$ is a positionally
sensitive probe for the local dynamics, covering a frequency range of
$\sim 15$ cm$^{-1}$. This is consistent with findings from selective
replacements of amino acids in PDZ2 which reported a frequency span of
$\sim 10$ cm$^{-1}$ for replacements of Val, Ala, or Glu by
AHA.\cite{hamm:2012} Furthermore, the long-time decay constants
$\tau_2$ range from $\sim 1$ to $\sim 10$ ps which compares with
experimentally measured correlation times of 3 ps.\cite{hamm:2012}
Attaching azide to alanine residues can yield dynamics that decays to
zero on the few ps time scale (i.e. $\Delta_0 \sim 0$ ps$^{-1}$) or to
a remaining inhomogeneous contribution of $\sim 0.5$ ps$^{-1}$
(corresponding to 2.5 cm$^{-1}$). One exciting prospect of this is to
determine how the spectroscopy and dynamics of the modification site
changes upon ligand binding to the active site for Lysozyme or other
proteins.\\

\section*{Acknowledgments}
The authors gratefully acknowledge financial support from the Swiss
National Science Foundation through grant 200021-117810 and to the
NCCR-MUST. The authors thank Prof. P. Hamm and Dr. D. Koner for
discussions on the experiments and some of the electronic structure
calculations. \\

\section*{Data Availability Statement}
The data that support the findings of this study are available from
the corresponding author upon reasonable request.

\newpage

\bibliography{lyso}

\providecommand{\latin}[1]{#1}
\makeatletter
\providecommand{\doi}
  {\begingroup\let\do\@makeother\dospecials
  \catcode`\{=1 \catcode`\}=2 \doi@aux}
\providecommand{\doi@aux}[1]{\endgroup\texttt{#1}}
\makeatother
\providecommand*\mcitethebibliography{\thebibliography}
\csname @ifundefined\endcsname{endmcitethebibliography}
  {\let\endmcitethebibliography\endthebibliography}{}
\begin{mcitethebibliography}{68}
\providecommand*\natexlab[1]{#1}
\providecommand*\mciteSetBstSublistMode[1]{}
\providecommand*\mciteSetBstMaxWidthForm[2]{}
\providecommand*\mciteBstWouldAddEndPuncttrue
  {\def\EndOfBibitem{\unskip.}}
\providecommand*\mciteBstWouldAddEndPunctfalse
  {\let\EndOfBibitem\relax}
\providecommand*\mciteSetBstMidEndSepPunct[3]{}
\providecommand*\mciteSetBstSublistLabelBeginEnd[3]{}
\providecommand*\EndOfBibitem{}
\mciteSetBstSublistMode{f}
\mciteSetBstMaxWidthForm{subitem}{(\alph{mcitesubitemcount})}
\mciteSetBstSublistLabelBeginEnd
  {\mcitemaxwidthsubitemform\space}
  {\relax}
  {\relax}

\bibitem[Plitzko \latin{et~al.}({2017})Plitzko, Schuler, and
  Selenko]{schuler:2017}
Plitzko,~J.~M.; Schuler,~B.; Selenko,~P. {Structural Biology outside the box -
  inside the cell}. \emph{Curr. Op. Struct. Biol.} \textbf{{2017}},
  \emph{{46}}, {110--121}\relax
\mciteBstWouldAddEndPuncttrue
\mciteSetBstMidEndSepPunct{\mcitedefaultmidpunct}
{\mcitedefaultendpunct}{\mcitedefaultseppunct}\relax
\EndOfBibitem
\bibitem[Guo and Zhou({2016})Guo, and Zhou]{zhou:2016}
Guo,~J.; Zhou,~H.-X. {Protein Allostery and Conformational Dynamics}.
  \emph{Chem. Rev.} \textbf{{2016}}, \emph{{116}}, {6503--6515}\relax
\mciteBstWouldAddEndPuncttrue
\mciteSetBstMidEndSepPunct{\mcitedefaultmidpunct}
{\mcitedefaultendpunct}{\mcitedefaultseppunct}\relax
\EndOfBibitem
\bibitem[Lu \latin{et~al.}({2019})Lu, He, Ni, and Zhang]{zhang:2019}
Lu,~S.; He,~X.; Ni,~D.; Zhang,~J. {Allosteric Modulator Discovery: From
  Serendipity to Structure-Based Design}. \emph{J. Med. Chem.} \textbf{{2019}},
  \emph{{62}}, {6405--6421}\relax
\mciteBstWouldAddEndPuncttrue
\mciteSetBstMidEndSepPunct{\mcitedefaultmidpunct}
{\mcitedefaultendpunct}{\mcitedefaultseppunct}\relax
\EndOfBibitem
\bibitem[Hamm and Zanni(2011)Hamm, and Zanni]{2DIRbook-Hamm-2011}
Hamm,~P.; Zanni,~M. \emph{{Concepts and Methods of 2D Infrared Spectroscopy}};
  Cambridge University Press: New York, 2011\relax
\mciteBstWouldAddEndPuncttrue
\mciteSetBstMidEndSepPunct{\mcitedefaultmidpunct}
{\mcitedefaultendpunct}{\mcitedefaultseppunct}\relax
\EndOfBibitem
\bibitem[Waegele \latin{et~al.}({2011})Waegele, Culik, and Gai]{gai:2011}
Waegele,~M.~M.; Culik,~R.~M.; Gai,~F. {Site-Specific Spectroscopic Reporters of
  the Local Electric Field, Hydration, Structure, and Dynamics of
  Biomolecules}. \emph{J. Phys. Chem. Lett.} \textbf{{2011}}, \emph{{2}},
  {2598--2609}\relax
\mciteBstWouldAddEndPuncttrue
\mciteSetBstMidEndSepPunct{\mcitedefaultmidpunct}
{\mcitedefaultendpunct}{\mcitedefaultseppunct}\relax
\EndOfBibitem
\bibitem[Koziol \latin{et~al.}({2015})Koziol, Johnson, Stucki-Buchli, Waldauer,
  and Hamm]{hamm.rev:2015}
Koziol,~K.~L.; Johnson,~P. J.~M.; Stucki-Buchli,~B.; Waldauer,~S.~A.; Hamm,~P.
  {Fast infrared spectroscopy of protein dynamics: advancing sensitivity and
  selectivity}. \emph{Curr. Op. Struct. Biol.} \textbf{{2015}}, \emph{{34}},
  {1--6}\relax
\mciteBstWouldAddEndPuncttrue
\mciteSetBstMidEndSepPunct{\mcitedefaultmidpunct}
{\mcitedefaultendpunct}{\mcitedefaultseppunct}\relax
\EndOfBibitem
\bibitem[Horness \latin{et~al.}({2015})Horness, Basom, and
  Thielges]{thielges:2015}
Horness,~R.~E.; Basom,~E.~J.; Thielges,~M.~C. {Site-selective characterization
  of Src homology 3 domain molecular recognition with cyanophenylalanine
  infrared probes}. \emph{Anal. Chem.} \textbf{{2015}}, \emph{{7}},
  {7234--7241}\relax
\mciteBstWouldAddEndPuncttrue
\mciteSetBstMidEndSepPunct{\mcitedefaultmidpunct}
{\mcitedefaultendpunct}{\mcitedefaultseppunct}\relax
\EndOfBibitem
\bibitem[Getahun \latin{et~al.}({2003})Getahun, Huang, Wang, De~Leon, DeGrado,
  and Gai]{gai:2003}
Getahun,~Z.; Huang,~C.; Wang,~T.; De~Leon,~B.; DeGrado,~W.; Gai,~F. {Using
  nitrile-derivatized amino acids as infrared probes of local environment}.
  \emph{J. Am. Chem. Soc.} \textbf{{2003}}, \emph{{125}}, {405--411}\relax
\mciteBstWouldAddEndPuncttrue
\mciteSetBstMidEndSepPunct{\mcitedefaultmidpunct}
{\mcitedefaultendpunct}{\mcitedefaultseppunct}\relax
\EndOfBibitem
\bibitem[Kozinski \latin{et~al.}({2008})Kozinski, Garrett-Roe, and
  Hamm]{hamm:2008}
Kozinski,~M.; Garrett-Roe,~S.; Hamm,~P. {2D-IR spectroscopy of the sulfhydryl
  band of cysteines in the hydrophobic core of proteins}. \emph{J. Phys. Chem.
  B} \textbf{{2008}}, \emph{{112}}, {7645--7650}\relax
\mciteBstWouldAddEndPuncttrue
\mciteSetBstMidEndSepPunct{\mcitedefaultmidpunct}
{\mcitedefaultendpunct}{\mcitedefaultseppunct}\relax
\EndOfBibitem
\bibitem[Zimmermann \latin{et~al.}({2011})Zimmermann, Thielges, Yu, Dawson, and
  Romesberg]{romesberg:2011}
Zimmermann,~J.; Thielges,~M.~C.; Yu,~W.; Dawson,~P.~E.; Romesberg,~F.~E.
  {Carbon-Deuterium Bonds as Site-Specific and Nonperturbative Probes for
  Time-Resolved Studies of Protein Dynamics and Folding}. \emph{J. Phys. Chem.
  Lett.} \textbf{{2011}}, \emph{{2}}, {412--416}\relax
\mciteBstWouldAddEndPuncttrue
\mciteSetBstMidEndSepPunct{\mcitedefaultmidpunct}
{\mcitedefaultendpunct}{\mcitedefaultseppunct}\relax
\EndOfBibitem
\bibitem[Woys \latin{et~al.}({2013})Woys, Mukherjee, Skoff, Moran, and
  Zanni]{zanni:2013}
Woys,~A.~M.; Mukherjee,~S.~S.; Skoff,~D.~R.; Moran,~S.~D.; Zanni,~M.~T. {A
  Strongly Absorbing Class of Non-Natural Labels for Probing Protein
  Electrostatics and Solvation with FTIR and 2D IR Spectroscopies}. \emph{J.
  Phys. Chem. B} \textbf{{2013}}, \emph{{117}}, {5009--5018}\relax
\mciteBstWouldAddEndPuncttrue
\mciteSetBstMidEndSepPunct{\mcitedefaultmidpunct}
{\mcitedefaultendpunct}{\mcitedefaultseppunct}\relax
\EndOfBibitem
\bibitem[Bagchi \latin{et~al.}({2012})Bagchi, Boxer, and
  Fayer]{fayer.ribo:2012}
Bagchi,~S.; Boxer,~S.~G.; Fayer,~M.~D. {Ribonuclease S Dynamics Measured Using
  a Nitrile Label with 2D IR Vibrational Echo Spectroscopy}. \emph{J. Phys.
  Chem. B} \textbf{{2012}}, \emph{{116}}, {4034--4042}\relax
\mciteBstWouldAddEndPuncttrue
\mciteSetBstMidEndSepPunct{\mcitedefaultmidpunct}
{\mcitedefaultendpunct}{\mcitedefaultseppunct}\relax
\EndOfBibitem
\bibitem[Zimmermann \latin{et~al.}({2011})Zimmermann, Thielges, Seo, Dawson,
  and Romesberg]{romesberg.cn:2011}
Zimmermann,~J.; Thielges,~M.~C.; Seo,~Y.~J.; Dawson,~P.~E.; Romesberg,~F.~E.
  {Cyano Groups as Probes of Protein Microenvironments and Dynamics}.
  \emph{Angew. Chem. Int. Ed.} \textbf{{2011}}, \emph{{50}}, {8333--8337}\relax
\mciteBstWouldAddEndPuncttrue
\mciteSetBstMidEndSepPunct{\mcitedefaultmidpunct}
{\mcitedefaultendpunct}{\mcitedefaultseppunct}\relax
\EndOfBibitem
\bibitem[van Wilderen \latin{et~al.}({2014})van Wilderen, Kern-Michler,
  Mueller-Werkmeister, and Bredenbeck]{bredenbeck:2014}
van Wilderen,~L. J. G.~W.; Kern-Michler,~D.; Mueller-Werkmeister,~H.~M.;
  Bredenbeck,~J. {Vibrational dynamics and solvatochromism of the label SCN in
  various solvents and hemoglobin by time dependent IR and 2D-IR spectroscopy}.
  \emph{Phys. Chem. Chem. Phys.} \textbf{{2014}}, \emph{{16}},
  {19643--19653}\relax
\mciteBstWouldAddEndPuncttrue
\mciteSetBstMidEndSepPunct{\mcitedefaultmidpunct}
{\mcitedefaultendpunct}{\mcitedefaultseppunct}\relax
\EndOfBibitem
\bibitem[Lee \latin{et~al.}({2018})Lee, Kossowska, Lim, Kim, Han, Kwak, and
  Cho]{cho:2018}
Lee,~G.; Kossowska,~D.; Lim,~J.; Kim,~S.; Han,~H.; Kwak,~K.; Cho,~M. {Cyanamide
  as an Infrared Reporter: Comparison of Vibrational Properties between
  Nitriles Bonded to N and C Atoms}. \emph{J. Phys. Chem. B} \textbf{{2018}},
  \emph{{122}}, {4035--4044}\relax
\mciteBstWouldAddEndPuncttrue
\mciteSetBstMidEndSepPunct{\mcitedefaultmidpunct}
{\mcitedefaultendpunct}{\mcitedefaultseppunct}\relax
\EndOfBibitem
\bibitem[Bloem \latin{et~al.}({2012})Bloem, Koziol, Waldauer, Buchli, Walser,
  Samatanga, Jelesarov, and Hamm]{hamm:2012}
Bloem,~R.; Koziol,~K.; Waldauer,~S.~A.; Buchli,~B.; Walser,~R.; Samatanga,~B.;
  Jelesarov,~I.; Hamm,~P. {Ligand Binding Studied by 2D IR Spectroscopy Using
  the Azidohomoalanine Label}. \emph{J. Phys. Chem. B} \textbf{{2012}},
  \emph{{116}}, {13705--13712}\relax
\mciteBstWouldAddEndPuncttrue
\mciteSetBstMidEndSepPunct{\mcitedefaultmidpunct}
{\mcitedefaultendpunct}{\mcitedefaultseppunct}\relax
\EndOfBibitem
\bibitem[Zanobini \latin{et~al.}({2018})Zanobini, Bozovic, Jankovic, Koziol,
  Johnson, Hamm, Gulzar, Wolf, and Stock]{stock:2018}
Zanobini,~C.; Bozovic,~O.; Jankovic,~B.; Koziol,~K.~L.; Johnson,~P. J.~M.;
  Hamm,~P.; Gulzar,~A.; Wolf,~S.; Stock,~G. {Azidohomoalanine: A Minimally
  Invasive, Versatile, and Sensitive Infrared Label in Proteins To Study Ligand
  Binding}. \emph{J. Phys. Chem. B} \textbf{{2018}}, \emph{{122}},
  {10118--10125}\relax
\mciteBstWouldAddEndPuncttrue
\mciteSetBstMidEndSepPunct{\mcitedefaultmidpunct}
{\mcitedefaultendpunct}{\mcitedefaultseppunct}\relax
\EndOfBibitem
\bibitem[Kiick \latin{et~al.}({2002})Kiick, Saxon, Tirrell, and
  Bertozzi]{bertozzi:2002}
Kiick,~K.; Saxon,~E.; Tirrell,~D.; Bertozzi,~C. {Incorporation of azides into
  recombinant proteins for chemoselective modification by the Staudinger
  ligation}. \emph{Proc. Natl. Acad. Sci.} \textbf{{2002}}, \emph{{99}},
  {19--24}\relax
\mciteBstWouldAddEndPuncttrue
\mciteSetBstMidEndSepPunct{\mcitedefaultmidpunct}
{\mcitedefaultendpunct}{\mcitedefaultseppunct}\relax
\EndOfBibitem
\bibitem[Koziol \latin{et~al.}({2015})Koziol, Johnson, Stucki-Buchli, Waldauer,
  and Hamm]{hamm:2015}
Koziol,~K.~L.; Johnson,~P. J.~M.; Stucki-Buchli,~B.; Waldauer,~S.~A.; Hamm,~P.
  {Fast Infrared Spectroscopy of Protein Dynamics: Advancing Sensitivity and
  Selectivity}. \emph{Curr. Op. Struct. Biol.} \textbf{{2015}}, \emph{{34}},
  {1--6}\relax
\mciteBstWouldAddEndPuncttrue
\mciteSetBstMidEndSepPunct{\mcitedefaultmidpunct}
{\mcitedefaultendpunct}{\mcitedefaultseppunct}\relax
\EndOfBibitem
\bibitem[Suydam \latin{et~al.}(2006)Suydam, Snow, Pande, and
  Boxer]{Suydam.halr.sci.2006}
Suydam,~I.~T.; Snow,~C.~D.; Pande,~V.~S.; Boxer,~S.~G. Electric Fields at the
  Active Site of an Enzyme : Direct Comparison of Experiment with Theory.
  \emph{Science} \textbf{2006}, \emph{313}, 200--204\relax
\mciteBstWouldAddEndPuncttrue
\mciteSetBstMidEndSepPunct{\mcitedefaultmidpunct}
{\mcitedefaultendpunct}{\mcitedefaultseppunct}\relax
\EndOfBibitem
\bibitem[Mondal and Meuwly(2017)Mondal, and Meuwly]{MM.lys:2017}
Mondal,~P.; Meuwly,~M. Vibrational Stark Spectroscopy for Assessing
  Ligand-Binding Strengths in a Protein. \emph{Phys. Chem. Chem. Phys.}
  \textbf{2017}, \emph{19}, 16131--16143\relax
\mciteBstWouldAddEndPuncttrue
\mciteSetBstMidEndSepPunct{\mcitedefaultmidpunct}
{\mcitedefaultendpunct}{\mcitedefaultseppunct}\relax
\EndOfBibitem
\bibitem[Salehi \latin{et~al.}({2019})Salehi, Koner, and
  Meuwly]{Salehi:N3-JPCB2019}
Salehi,~S.~M.; Koner,~D.; Meuwly,~M. {Vibrational Spectroscopy of N$_{3}^{-}$
  in the Gas and Condensed Phase}. \emph{J. Phys. Chem. B} \textbf{{2019}},
  \emph{{123}}, {3282--3290}\relax
\mciteBstWouldAddEndPuncttrue
\mciteSetBstMidEndSepPunct{\mcitedefaultmidpunct}
{\mcitedefaultendpunct}{\mcitedefaultseppunct}\relax
\EndOfBibitem
\bibitem[Ho and Rabitz(1996)Ho, and Rabitz]{RKHS-Rabitz-1996}
Ho,~T.-S.; Rabitz,~R. A General Method for Constructing Multidimensional
  Molecular Potential Energy Surfaces from ab Initio Calculations. \emph{J.
  Chem. Phys.} \textbf{1996}, \emph{104}, 2584--2597\relax
\mciteBstWouldAddEndPuncttrue
\mciteSetBstMidEndSepPunct{\mcitedefaultmidpunct}
{\mcitedefaultendpunct}{\mcitedefaultseppunct}\relax
\EndOfBibitem
\bibitem[Unke and Meuwly(2017)Unke, and Meuwly]{MM.rkhs:2017}
Unke,~O.~T.; Meuwly,~M. Toolkit for the Construction of Reproducing
  Kernel-Based Representations of Data: Application to Multidimensional
  Potential Energy Surfaces. \emph{J. Chem. Inf. Model.} \textbf{2017},
  \emph{57}, 1923--1931\relax
\mciteBstWouldAddEndPuncttrue
\mciteSetBstMidEndSepPunct{\mcitedefaultmidpunct}
{\mcitedefaultendpunct}{\mcitedefaultseppunct}\relax
\EndOfBibitem
\bibitem[Brooks \latin{et~al.}(2009)Brooks, Brooks~III, MacKerell~Jr., Nilsson,
  Petrella, Roux, Won, Archontis, Bartels, Boresch, Caflisch, Caves, Cui,
  Dinner, Feig, Fischer, Gao, Hodoscek, Im, Kuczera, Lazaridis, Ma,
  Ovchinnikov, Paci, Pastor, Post, Schaefer, Tidor, Venable, Woodcock, Wu,
  Yang, York, and Karplus]{Charmm-Brooks-2009}
Brooks,~B.~R.; Brooks~III,~C.~L.; MacKerell~Jr.,~A.~D.; Nilsson,~L.;
  Petrella,~R.~J.; Roux,~B.; Won,~Y.; Archontis,~G.; Bartels,~C.; Boresch,~S.
  \latin{et~al.}  CHARMM: The Biomolecular Simulation Program. \emph{J. Comp.
  Chem.} \textbf{2009}, \emph{30}, 1545--1614\relax
\mciteBstWouldAddEndPuncttrue
\mciteSetBstMidEndSepPunct{\mcitedefaultmidpunct}
{\mcitedefaultendpunct}{\mcitedefaultseppunct}\relax
\EndOfBibitem
\bibitem[MacKerell \latin{et~al.}(1998)MacKerell, Bashford, Bellott, Dunbrack,
  Evanseck, Field, Fischer, Gao, Guo, Ha, Joseph-McCarthy, Kuchnir, Kuczera,
  Lau, Mattos, Michnick, Ngo, Nguyen, Prodhom, Reiher, Roux, Schlenkrich,
  Smith, Stote, Straub, Watanabe, Wirkiewicz-Kuczera, Yin, and
  Karplus]{charmmFF22}
MacKerell,~A.~D.; Bashford,~D.; Bellott,~M.; Dunbrack,~R.~L.; Evanseck,~J.~D.;
  Field,~M.~J.; Fischer,~S.; Gao,~J.; Guo,~H.; Ha,~S. \latin{et~al.}  All-atom
  Empirical Potential for Molecular Modeling and Dynamics Studies of Proteins.
  \emph{J. Phys. Chem. B} \textbf{1998}, \emph{102}, 3586--3616\relax
\mciteBstWouldAddEndPuncttrue
\mciteSetBstMidEndSepPunct{\mcitedefaultmidpunct}
{\mcitedefaultendpunct}{\mcitedefaultseppunct}\relax
\EndOfBibitem
\bibitem[Chiba-Kamoshida \latin{et~al.}()Chiba-Kamoshida, Matsui, Ostermann,
  Chatake, Ohhara, Tanaka, Yutani, and Niimura]{pdblyso-2009}
Chiba-Kamoshida,~K.; Matsui,~T.; Ostermann,~A.; Chatake,~T.; Ohhara,~T.;
  Tanaka,~I.; Yutani,~K.; Niimura,~N. X-ray crystal structure of wild type
  human lysozyme in D$_2$O. DOI: 10.2210/pdb3fe0/pdb\relax
\mciteBstWouldAddEndPuncttrue
\mciteSetBstMidEndSepPunct{\mcitedefaultmidpunct}
{\mcitedefaultendpunct}{\mcitedefaultseppunct}\relax
\EndOfBibitem
\bibitem[Jorgensen \latin{et~al.}(1983)Jorgensen, Chandrasekhar, Madura, Impey,
  and Klein]{TIP3P-Jorgensen-1983}
Jorgensen,~W.~L.; Chandrasekhar,~J.; Madura,~J.~D.; Impey,~R.~W.; Klein,~M.~L.
  Comparison of Simple Potential Functions for Simulating Liquid Water.
  \emph{J. Chem. Phys.} \textbf{1983}, \emph{79}, 926--935\relax
\mciteBstWouldAddEndPuncttrue
\mciteSetBstMidEndSepPunct{\mcitedefaultmidpunct}
{\mcitedefaultendpunct}{\mcitedefaultseppunct}\relax
\EndOfBibitem
\bibitem[Steinbach and Brooks(1994)Steinbach, and Brooks]{Steinbach1994}
Steinbach,~P.~J.; Brooks,~B.~R. New Spherical-Cutoff Methods for Long-Range
  Forces in Macromolecular Simulation. \emph{J. Comput. Chem.} \textbf{1994},
  \emph{15}, 667--683\relax
\mciteBstWouldAddEndPuncttrue
\mciteSetBstMidEndSepPunct{\mcitedefaultmidpunct}
{\mcitedefaultendpunct}{\mcitedefaultseppunct}\relax
\EndOfBibitem
\bibitem[Gunsteren and Berendsen(1997)Gunsteren, and
  Berendsen]{SHAKE-Gunsteren-1997}
Gunsteren,~W.~V.; Berendsen,~H. Algorithms for Macromolecular Dynamics and
  Constraint Dynamics. \emph{Mol. Phys.} \textbf{1997}, \emph{34},
  1311--1327\relax
\mciteBstWouldAddEndPuncttrue
\mciteSetBstMidEndSepPunct{\mcitedefaultmidpunct}
{\mcitedefaultendpunct}{\mcitedefaultseppunct}\relax
\EndOfBibitem
\bibitem[Schwilk \latin{et~al.}({2017})Schwilk, Ma, Koeppl, and
  Werner]{lccsd-schwilk-2017}
Schwilk,~M.; Ma,~Q.; Koeppl,~C.; Werner,~H.-J. {Scalable Electron Correlation
  Methods. 3. Efficient and Accurate Parallel Local Coupled Cluster with Pair
  Natural Orbitals (PNO-LCCSD)}. \emph{J. Chem. Theo. Comp.} \textbf{{2017}},
  \emph{{13}}, {3650--3675}\relax
\mciteBstWouldAddEndPuncttrue
\mciteSetBstMidEndSepPunct{\mcitedefaultmidpunct}
{\mcitedefaultendpunct}{\mcitedefaultseppunct}\relax
\EndOfBibitem
\bibitem[Ma \latin{et~al.}({2018})Ma, Schwilk, Koeppl, and
  Werner]{lccsdf12-schwilk-2017}
Ma,~Q.; Schwilk,~M.; Koeppl,~C.; Werner,~H.-J. {Scalable Electron Correlation
  Methods. 4. Parallel Explicitly Correlated Local Coupled Cluster with Pair
  Natural Orbitals (PNO-LCCSD-F12) (vol 13, pg 4871, 2017)}. \emph{J. Chem.
  Theo. Comp.} \textbf{{2018}}, \emph{{14}}, {6750}\relax
\mciteBstWouldAddEndPuncttrue
\mciteSetBstMidEndSepPunct{\mcitedefaultmidpunct}
{\mcitedefaultendpunct}{\mcitedefaultseppunct}\relax
\EndOfBibitem
\bibitem[Dunning(1989)]{DUN:JCP89}
Dunning,~T.~H.,~Jr. Gaussian basis sets for use in correlated molecular
  calculations. I. The atoms boron through neon and hydrogen. \emph{J. Chem.
  Phys.} \textbf{1989}, \emph{90}, 1007--1023\relax
\mciteBstWouldAddEndPuncttrue
\mciteSetBstMidEndSepPunct{\mcitedefaultmidpunct}
{\mcitedefaultendpunct}{\mcitedefaultseppunct}\relax
\EndOfBibitem
\bibitem[Werner \latin{et~al.}(2012)Werner, Knowles, Knizia, Manby,
  {Sch\"{u}tz}, Celani, Gy\"orffy, Kats, Korona, Lindh, Mitrushenkov, Rauhut,
  Shamasundar, Adler, Amos, Bennie, Bernhardsson, Berning, Cooper, Deegan,
  Dobbyn, Eckert, Goll, Hampel, Hesselmann, Hetzer, Hrenar, Jansen, K\"oppl,
  Lee, Liu, Lloyd, Ma, Mata, May, McNicholas, Meyer, {Miller III}, Mura,
  Nicklass, O'Neill, Palmieri, Peng, Pfl\"uger, Pitzer, Reiher, Shiozaki,
  Stoll, Stone, Tarroni, Thorsteinsson, and Wang]{molpro2}
Werner,~H.-J.; Knowles,~P.~J.; Knizia,~G.; Manby,~F.~R.; {Sch\"{u}tz},~M.;
  Celani,~P.; Gy\"orffy,~W.; Kats,~D.; Korona,~T.; Lindh,~R. \latin{et~al.}
  MOLPRO, Version 2012.1, A Package of ab Initio Programs. 2012\relax
\mciteBstWouldAddEndPuncttrue
\mciteSetBstMidEndSepPunct{\mcitedefaultmidpunct}
{\mcitedefaultendpunct}{\mcitedefaultseppunct}\relax
\EndOfBibitem
\bibitem[Zoete \latin{et~al.}(2011)Zoete, Cuendet, Grosdidier, and
  Michielin]{swissparam-Zoete-2011}
Zoete,~V.; Cuendet,~M.; Grosdidier,~A.; Michielin,~O. SwissParam: A Fast Force
  Field Generation Tool for Small Organic Molecules. \emph{J. Comp. Chem.}
  \textbf{2011}, \emph{32}, 2359--2368\relax
\mciteBstWouldAddEndPuncttrue
\mciteSetBstMidEndSepPunct{\mcitedefaultmidpunct}
{\mcitedefaultendpunct}{\mcitedefaultseppunct}\relax
\EndOfBibitem
\bibitem[Meuwly and Hutson({1999})Meuwly, and Hutson]{MM.heh2.1998}
Meuwly,~M.; Hutson,~J. {The Potential Energy Surface and Near-dissociation
  States of He-H$_2^+$}. \emph{J. Chem. Phys.} \textbf{{1999}}, \emph{{110}},
  {3418--3427}\relax
\mciteBstWouldAddEndPuncttrue
\mciteSetBstMidEndSepPunct{\mcitedefaultmidpunct}
{\mcitedefaultendpunct}{\mcitedefaultseppunct}\relax
\EndOfBibitem
\bibitem[Sold{\'a}n and Hutson(2000)Sold{\'a}n, and Hutson]{hutson:2000}
Sold{\'a}n,~P.; Hutson,~J.~M. On the long-range and short-range behavior of
  potentials from reproducing kernel Hilbert space interpolation. \emph{J.
  Chem. Phys.} \textbf{2000}, \emph{112}, 4415--4416\relax
\mciteBstWouldAddEndPuncttrue
\mciteSetBstMidEndSepPunct{\mcitedefaultmidpunct}
{\mcitedefaultendpunct}{\mcitedefaultseppunct}\relax
\EndOfBibitem
\bibitem[Hollebeek \latin{et~al.}(1999)Hollebeek, Ho, and
  Rabitz]{hollebeek1999constructing}
Hollebeek,~T.; Ho,~T.-S.; Rabitz,~H. Constructing Multidimensional Molecular
  Potential Energy Surfaces from ab Initio Data. \emph{Ann. Rev. Phys. Chem.}
  \textbf{1999}, \emph{50}, 537--570\relax
\mciteBstWouldAddEndPuncttrue
\mciteSetBstMidEndSepPunct{\mcitedefaultmidpunct}
{\mcitedefaultendpunct}{\mcitedefaultseppunct}\relax
\EndOfBibitem
\bibitem[Foster and Weinhold(1980)Foster, and Weinhold]{NBO-Foster-1980}
Foster,~J.~P.; Weinhold,~F. Natural Hybrid Orbitals. \emph{J. Am. Chem. Soc.}
  \textbf{1980}, \emph{102}, 7211--18\relax
\mciteBstWouldAddEndPuncttrue
\mciteSetBstMidEndSepPunct{\mcitedefaultmidpunct}
{\mcitedefaultendpunct}{\mcitedefaultseppunct}\relax
\EndOfBibitem
\bibitem[Frisch \latin{et~al.}(2016)Frisch, Trucks, Schlegel, Scuseria, Robb,
  Cheeseman, Scalmani, Barone, Petersson, Nakatsuji, Li, Caricato, Marenich,
  Bloino, Janesko, Gomperts, Mennucci, Hratchian, Ortiz, Izmaylov, Sonnenberg,
  Williams-Young, Ding, Lipparini, Egidi, Goings, Peng, Petrone, Henderson,
  Ranasinghe, Zakrzewski, Gao, Rega, Zheng, Liang, Hada, Ehara, Toyota, Fukuda,
  Hasegawa, Ishida, Nakajima, Honda, Kitao, Nakai, Vreven, Throssell,
  Montgomery, Peralta, Ogliaro, Bearpark, Heyd, Brothers, Kudin, Staroverov,
  Keith, Kobayashi, Normand, Raghavachari, Rendell, Burant, Iyengar, Tomasi,
  Cossi, Millam, Klene, Adamo, Cammi, Ochterski, Martin, Morokuma, Farkas,
  Foresman, and Fox]{g09}
Frisch,~M.~J.; Trucks,~G.~W.; Schlegel,~H.~B.; Scuseria,~G.~E.; Robb,~M.~A.;
  Cheeseman,~J.~R.; Scalmani,~G.; Barone,~V.; Petersson,~G.~A.; Nakatsuji,~H.
  \latin{et~al.}  Gaussian˜16 {R}evision {C}.09. 2016; Gaussian Inc.
  Wallingford CT\relax
\mciteBstWouldAddEndPuncttrue
\mciteSetBstMidEndSepPunct{\mcitedefaultmidpunct}
{\mcitedefaultendpunct}{\mcitedefaultseppunct}\relax
\EndOfBibitem
\bibitem[Salehi \latin{et~al.}({2020})Salehi, Koner, and
  Meuwly]{MM.insulin:2020}
Salehi,~S.~M.; Koner,~D.; Meuwly,~M. {Dynamics and Infrared Spectrocopy of
  Monomeric and Dimeric Wild Type and Mutant Insulin}. \emph{J. Phys. Chem. B}
  \textbf{{2020}}, \emph{{in print}}, {in print}\relax
\mciteBstWouldAddEndPuncttrue
\mciteSetBstMidEndSepPunct{\mcitedefaultmidpunct}
{\mcitedefaultendpunct}{\mcitedefaultseppunct}\relax
\EndOfBibitem
\bibitem[Maekawa \latin{et~al.}({2004})Maekawa, Ohta, and
  Tominaga]{n3exph20-Maekawa-2004}
Maekawa,~H.; Ohta,~K.; Tominaga,~K. {Spectral Diffusion of the Anti-Symmetric
  Stretching Mode of Azide Ion in a Reverse Micelle Studied by Infrared
  Three-Pulse Photon Echo Method}. \emph{Phys. Chem. Chem. Phys.}
  \textbf{{2004}}, \emph{{6}}, {4074--4077}\relax
\mciteBstWouldAddEndPuncttrue
\mciteSetBstMidEndSepPunct{\mcitedefaultmidpunct}
{\mcitedefaultendpunct}{\mcitedefaultseppunct}\relax
\EndOfBibitem
\bibitem[Zhong \latin{et~al.}({2003})Zhong, Baronavski, and
  Owrutsky]{timedecay-Owrutsky-2003}
Zhong,~Q.; Baronavski,~A.; Owrutsky,~J. {Vibrational Energy Relaxation of
  Aqueous Azide Ion Confined in Reverse Micelles}. \emph{{J. Chem. Phys.}}
  \textbf{{2003}}, \emph{{118}}, {7074--7080}\relax
\mciteBstWouldAddEndPuncttrue
\mciteSetBstMidEndSepPunct{\mcitedefaultmidpunct}
{\mcitedefaultendpunct}{\mcitedefaultseppunct}\relax
\EndOfBibitem
\bibitem[Moller \latin{et~al.}({2004})Moller, Rey, and Hynes]{hynes:2004}
Moller,~K.; Rey,~R.; Hynes,~J. {Hydrogen Bond Dynamics in Water and Ultrafast
  Infrared Spectroscopy: A Theoretical Study}. \emph{J. Phys. Chem. A}
  \textbf{{2004}}, \emph{{108}}, {1275--1289}\relax
\mciteBstWouldAddEndPuncttrue
\mciteSetBstMidEndSepPunct{\mcitedefaultmidpunct}
{\mcitedefaultendpunct}{\mcitedefaultseppunct}\relax
\EndOfBibitem
\bibitem[{Virtanen} \latin{et~al.}(2020){Virtanen}, {Gommers}, {Oliphant},
  {Haberland}, {Reddy}, {Cournapeau}, {Burovski}, {Peterson}, {Weckesser},
  {Bright}, {van der Walt}, {Brett}, {Wilson}, {Jarrod Millman}, {Mayorov},
  {Nelson}, {Jones}, {Kern}, {Larson}, {Carey}, {Polat}, {Feng}, {Moore}, {Vand
  erPlas}, {Laxalde}, {Perktold}, {Cimrman}, {Henriksen}, {Quintero}, {Harris},
  {Archibald}, {Ribeiro}, {Pedregosa}, {van Mulbregt}, and
  {Contributors}]{2020SciPy-NMeth}
{Virtanen},~P.; {Gommers},~R.; {Oliphant},~T.~E.; {Haberland},~M.; {Reddy},~T.;
  {Cournapeau},~D.; {Burovski},~E.; {Peterson},~P.; {Weckesser},~W.;
  {Bright},~J. \latin{et~al.}  {SciPy 1.0: Fundamental Algorithms for
  Scientific Computing in Python}. \emph{Nature Methods} \textbf{2020},
  \emph{17}, 261--272\relax
\mciteBstWouldAddEndPuncttrue
\mciteSetBstMidEndSepPunct{\mcitedefaultmidpunct}
{\mcitedefaultendpunct}{\mcitedefaultseppunct}\relax
\EndOfBibitem
\bibitem[Okuda \latin{et~al.}({2015})Okuda, Ohta, and Tominaga]{alan3-jcp-2015}
Okuda,~M.; Ohta,~K.; Tominaga,~K. {Vibrational dynamics of azide-derivatized
  amino acids studied by nonlinear infrared spectroscopy}. \emph{{jcp}}
  \textbf{{2015}}, \emph{{142}}\relax
\mciteBstWouldAddEndPuncttrue
\mciteSetBstMidEndSepPunct{\mcitedefaultmidpunct}
{\mcitedefaultendpunct}{\mcitedefaultseppunct}\relax
\EndOfBibitem
\bibitem[Bowman and Gazdy(1991)Bowman, and Gazdy]{JMB91morphing}
Bowman,~J.~M.; Gazdy,~B. A simple method to adjust potential energy surfaces:
  Application to HCO. \emph{J. Chem. Phys.} \textbf{1991}, \emph{94},
  816--817\relax
\mciteBstWouldAddEndPuncttrue
\mciteSetBstMidEndSepPunct{\mcitedefaultmidpunct}
{\mcitedefaultendpunct}{\mcitedefaultseppunct}\relax
\EndOfBibitem
\bibitem[Meuwly and Hutson(1999)Meuwly, and Hutson]{MM.morph:1999}
Meuwly,~M.; Hutson,~J.~M. Morphing ab Initio potentials: A systematic study of
  Ne--HF. \emph{J. Chem. Phys.} \textbf{1999}, \emph{110}, 8338--8347\relax
\mciteBstWouldAddEndPuncttrue
\mciteSetBstMidEndSepPunct{\mcitedefaultmidpunct}
{\mcitedefaultendpunct}{\mcitedefaultseppunct}\relax
\EndOfBibitem
\bibitem[Li \latin{et~al.}({2006})Li, Schmidt, Piryatinski, Lawrence, and
  Skinner]{skinner:2006}
Li,~S.; Schmidt,~J.~R.; Piryatinski,~A.; Lawrence,~C.~P.; Skinner,~J.~L.
  {Vibrational Spectral Diffusion of Azide in Water}. \emph{J. Phys. Chem. B}
  \textbf{{2006}}, \emph{{110}}, {18933--18938}\relax
\mciteBstWouldAddEndPuncttrue
\mciteSetBstMidEndSepPunct{\mcitedefaultmidpunct}
{\mcitedefaultendpunct}{\mcitedefaultseppunct}\relax
\EndOfBibitem
\bibitem[Lee \latin{et~al.}(2013)Lee, Carr, G\"{o}llner, Hamm, and
  Meuwly]{MM.cn:2013}
Lee,~M.~W.; Carr,~J.~K.; G\"{o}llner,~M.; Hamm,~P.; Meuwly,~M. 2D IR Spectra of
  Cyanide in Water Investigated by Molecular Dynamics Simulations. \emph{J.
  Chem. Phys.} \textbf{2013}, \emph{139}, 054506\relax
\mciteBstWouldAddEndPuncttrue
\mciteSetBstMidEndSepPunct{\mcitedefaultmidpunct}
{\mcitedefaultendpunct}{\mcitedefaultseppunct}\relax
\EndOfBibitem
\bibitem[Hamm \latin{et~al.}(1998)Hamm, Lim, and Hochstrasser]{hamm:1998}
Hamm,~P.; Lim,~M.; Hochstrasser,~R.~M. {Structure of the Amide I Band of
  Peptides Measured by Femtosecond Nonlinear-Infrared Spectroscopy}. \emph{J.
  Phys. Chem. B} \textbf{1998}, \emph{5647}, 6123--6138\relax
\mciteBstWouldAddEndPuncttrue
\mciteSetBstMidEndSepPunct{\mcitedefaultmidpunct}
{\mcitedefaultendpunct}{\mcitedefaultseppunct}\relax
\EndOfBibitem
\bibitem[Cazade \latin{et~al.}({2014})Cazade, Bereau, and
  Meuwly]{NMA2dir-MM-2014}
Cazade,~P.-A.; Bereau,~T.; Meuwly,~M. {Computational Two-Dimensional Infrared
  Spectroscopy without Maps: N-Methylacetamide in Water}. \emph{J. Phys. Chem.
  B} \textbf{{2014}}, \emph{{118}}, {8135--8147}\relax
\mciteBstWouldAddEndPuncttrue
\mciteSetBstMidEndSepPunct{\mcitedefaultmidpunct}
{\mcitedefaultendpunct}{\mcitedefaultseppunct}\relax
\EndOfBibitem
\bibitem[Koner \latin{et~al.}(2020)Koner, Salehi, Mondal, and
  Meuwly]{MM.rev.jcp:2020}
Koner,~D.; Salehi,~S.~M.; Mondal,~P.; Meuwly,~M. Non-conventional force fields
  for applications in spectroscopy and chemical reaction dynamics. \emph{J.
  Chem. Phys.} \textbf{2020}, \emph{153}, 010901\relax
\mciteBstWouldAddEndPuncttrue
\mciteSetBstMidEndSepPunct{\mcitedefaultmidpunct}
{\mcitedefaultendpunct}{\mcitedefaultseppunct}\relax
\EndOfBibitem
\bibitem[Shin and Willard({2018})Shin, and Willard]{Willard-jctc-2018}
Shin,~S.; Willard,~A.~P. {Characterizing Hydration Properties Based on the
  Orientational Structure of Interfacial Water Molecules}. \emph{J. Chem. Theo.
  Comp.} \textbf{{2018}}, \emph{{14}}, {461--465}\relax
\mciteBstWouldAddEndPuncttrue
\mciteSetBstMidEndSepPunct{\mcitedefaultmidpunct}
{\mcitedefaultendpunct}{\mcitedefaultseppunct}\relax
\EndOfBibitem
\bibitem[Shin and Willard({2018})Shin, and Willard]{Willard-jpcb-2018}
Shin,~S.; Willard,~A.~P. {Water's Interfacial Hydrogen Bonding Structure
  Reveals the Effective Strength of Surface-Water Interactions}. \emph{J. Phys.
  Chem. B} \textbf{{2018}}, \emph{{122}}, {6781--6789}\relax
\mciteBstWouldAddEndPuncttrue
\mciteSetBstMidEndSepPunct{\mcitedefaultmidpunct}
{\mcitedefaultendpunct}{\mcitedefaultseppunct}\relax
\EndOfBibitem
\bibitem[Pezzella \latin{et~al.}(2020)Pezzella, El~Hage, Niesen, Shin, Willard,
  Meuwly, and Karplus]{MM.hb:2020}
Pezzella,~M.; El~Hage,~K.; Niesen,~M.~J.; Shin,~S.; Willard,~A.~P.; Meuwly,~M.;
  Karplus,~M. Water dynamics around proteins: T-and R-States of hemoglobin and
  melittin. \emph{J. Phys. Chem. B} \textbf{2020}, \emph{124}, 6540--6554\relax
\mciteBstWouldAddEndPuncttrue
\mciteSetBstMidEndSepPunct{\mcitedefaultmidpunct}
{\mcitedefaultendpunct}{\mcitedefaultseppunct}\relax
\EndOfBibitem
\bibitem[Desmond \latin{et~al.}({2019})Desmond, Koner, and
  Meuwly]{MM.insulin:2019}
Desmond,~J.~L.; Koner,~D.; Meuwly,~M. {Probing the Differential Dynamics of the
  Monomeric and Dimeric Insulin from Amide-I IR Spectroscopy}. \emph{J. Phys.
  Chem. B} \textbf{{2019}}, \emph{{123}}, {6588--6598}\relax
\mciteBstWouldAddEndPuncttrue
\mciteSetBstMidEndSepPunct{\mcitedefaultmidpunct}
{\mcitedefaultendpunct}{\mcitedefaultseppunct}\relax
\EndOfBibitem
\bibitem[Baldwin \latin{et~al.}(1995)Baldwin, Bhat, Gulnik, Liu, Topol, Kiso,
  Mimoto, Mitsuya, and Erickson]{erickson:1995}
Baldwin,~E.~T.; Bhat,~T.~N.; Gulnik,~S.; Liu,~B.; Topol,~I.~A.; Kiso,~Y.;
  Mimoto,~T.; Mitsuya,~H.; Erickson,~J.~W. Structure of HIV-1 protease with
  KNI-272, a tight-binding transition-state analog containing
  allophenylnorstatine. \emph{Structure} \textbf{1995}, \emph{3},
  581--590\relax
\mciteBstWouldAddEndPuncttrue
\mciteSetBstMidEndSepPunct{\mcitedefaultmidpunct}
{\mcitedefaultendpunct}{\mcitedefaultseppunct}\relax
\EndOfBibitem
\bibitem[Prashar \latin{et~al.}(2009)Prashar, Bihani, Das, Ferrer, and
  Hosur]{prashar:2009}
Prashar,~V.; Bihani,~S.; Das,~A.; Ferrer,~J.-L.; Hosur,~M. Catalytic water
  co-existing with a product peptide in the active site of HIV-1 protease
  revealed by X-ray structure analysis. \emph{PloS one} \textbf{2009},
  \emph{4}, e7860\relax
\mciteBstWouldAddEndPuncttrue
\mciteSetBstMidEndSepPunct{\mcitedefaultmidpunct}
{\mcitedefaultendpunct}{\mcitedefaultseppunct}\relax
\EndOfBibitem
\bibitem[Raghunathan \latin{et~al.}({2018})Raghunathan, El~Hage, Desmond,
  Zhang, and Meuwly]{MM.insulin:2018}
Raghunathan,~S.; El~Hage,~K.; Desmond,~J.~L.; Zhang,~L.; Meuwly,~M. {The Role
  of Water in the Stability of Wild-type and Mutant Insulin Dimers}. \emph{J.
  Phys. Chem. B} \textbf{{2018}}, \emph{{122}}, {7038--7048}\relax
\mciteBstWouldAddEndPuncttrue
\mciteSetBstMidEndSepPunct{\mcitedefaultmidpunct}
{\mcitedefaultendpunct}{\mcitedefaultseppunct}\relax
\EndOfBibitem
\bibitem[Schiro \latin{et~al.}(2015)Schiro, Fichou, Gallat, Wood, Gabel,
  Moulin, Haertlein, Heyden, Colletier, Orecchini, Paciaroni, Wuttke, Tobias,
  and Weik]{weik:2015}
Schiro,~G.; Fichou,~Y.; Gallat,~F.-X.; Wood,~K.; Gabel,~F.; Moulin,~M.;
  Haertlein,~M.; Heyden,~M.; Colletier,~J.-P.; Orecchini,~A. \latin{et~al.}
  {Translational diffusion of hydration water correlates with functional
  motions in folded and intrinsically disordered proteins}. \emph{Nuovo Cim.}
  \textbf{2015}, \emph{6}, 1--8\relax
\mciteBstWouldAddEndPuncttrue
\mciteSetBstMidEndSepPunct{\mcitedefaultmidpunct}
{\mcitedefaultendpunct}{\mcitedefaultseppunct}\relax
\EndOfBibitem
\bibitem[Pocker(2000)]{pocker:2000}
Pocker,~Y. Water in enzyme reactions: biophysical aspects of
  hydration-dehydration processes. \emph{Cell. Mol. Life Sci.} \textbf{2000},
  \emph{57}, 1008--1017\relax
\mciteBstWouldAddEndPuncttrue
\mciteSetBstMidEndSepPunct{\mcitedefaultmidpunct}
{\mcitedefaultendpunct}{\mcitedefaultseppunct}\relax
\EndOfBibitem
\bibitem[Pal and Zewail({2004})Pal, and Zewail]{zewail:2004}
Pal,~S.; Zewail,~A. {Dynamics of water in biological recognition}. \emph{Chem.
  Rev.} \textbf{{2004}}, \emph{{104}}, {2099--2123}\relax
\mciteBstWouldAddEndPuncttrue
\mciteSetBstMidEndSepPunct{\mcitedefaultmidpunct}
{\mcitedefaultendpunct}{\mcitedefaultseppunct}\relax
\EndOfBibitem
\bibitem[Zhang \latin{et~al.}({2016})Zhang, Jones, Fitzpatrick, Peng, Feng,
  Baiz, and Tokmakoff]{tokmakoff:2016.2}
Zhang,~X.-X.; Jones,~K.~C.; Fitzpatrick,~A.; Peng,~C.~S.; Feng,~C.-J.;
  Baiz,~C.~R.; Tokmakoff,~A. {Studying Protein-Protein Binding through T-Jump
  Induced Dissociation: Transient 2D IR Spectroscopy of Insulin Dimer}.
  \emph{J. Phys. Chem. B} \textbf{{2016}}, \emph{{120}}, {5134--5145}\relax
\mciteBstWouldAddEndPuncttrue
\mciteSetBstMidEndSepPunct{\mcitedefaultmidpunct}
{\mcitedefaultendpunct}{\mcitedefaultseppunct}\relax
\EndOfBibitem
\bibitem[Zhang and Tokmakoff({2020})Zhang, and Tokmakoff]{tokmakoff:2020}
Zhang,~X.-X.; Tokmakoff,~A. {Revealing the Dynamical Role of Co-solvents in the
  Coupled Folding and Dimerization of Insulin}. \emph{J. Phys. Chem. Lett.}
  \textbf{{2020}}, \emph{{11}}, {4353--4358}\relax
\mciteBstWouldAddEndPuncttrue
\mciteSetBstMidEndSepPunct{\mcitedefaultmidpunct}
{\mcitedefaultendpunct}{\mcitedefaultseppunct}\relax
\EndOfBibitem
\bibitem[Antoszewski \latin{et~al.}({2020})Antoszewski, Feng, Vani, Thiede,
  Hong, Weare, Tokmakoff, and Dinner]{dinner:2020}
Antoszewski,~A.; Feng,~C.-J.; Vani,~B.~P.; Thiede,~E.~H.; Hong,~L.; Weare,~J.;
  Tokmakoff,~A.; Dinner,~A.~R. {Insulin Dissociates by Diverse Mechanisms of
  Coupled Unfolding and Unbinding}. \emph{J. Phys. Chem. B} \textbf{{2020}},
  \emph{{124}}, {5571--5587}\relax
\mciteBstWouldAddEndPuncttrue
\mciteSetBstMidEndSepPunct{\mcitedefaultmidpunct}
{\mcitedefaultendpunct}{\mcitedefaultseppunct}\relax
\EndOfBibitem
\bibitem[Miller \latin{et~al.}({2013})Miller, Bourassa, and Smith]{smith:2013}
Miller,~L.~M.; Bourassa,~M.~W.; Smith,~R.~J. {FTIR spectroscopic imaging of
  protein aggregation in living cells}. \emph{Bioch. Bioph. Acta}
  \textbf{{2013}}, \emph{{1828}}, {2339--2346}\relax
\mciteBstWouldAddEndPuncttrue
\mciteSetBstMidEndSepPunct{\mcitedefaultmidpunct}
{\mcitedefaultendpunct}{\mcitedefaultseppunct}\relax
\EndOfBibitem
\end{mcitethebibliography}


\providecommand{\latin}[1]{#1}
\makeatletter
\providecommand{\doi}
  {\begingroup\let\do\@makeother\dospecials
  \catcode`\{=1 \catcode`\}=2 \doi@aux}
\providecommand{\doi@aux}[1]{\endgroup\texttt{#1}}
\makeatother
\providecommand*\mcitethebibliography{\thebibliography}
\csname @ifundefined\endcsname{endmcitethebibliography}
  {\let\endmcitethebibliography\endthebibliography}{}
\begin{mcitethebibliography}{0}
\providecommand*\natexlab[1]{#1}
\providecommand*\mciteSetBstSublistMode[1]{}
\providecommand*\mciteSetBstMaxWidthForm[2]{}
\providecommand*\mciteBstWouldAddEndPuncttrue
  {\def\EndOfBibitem{\unskip.}}
\providecommand*\mciteBstWouldAddEndPunctfalse
  {\let\EndOfBibitem\relax}
\providecommand*\mciteSetBstMidEndSepPunct[3]{}
\providecommand*\mciteSetBstSublistLabelBeginEnd[3]{}
\providecommand*\EndOfBibitem{}
\mciteSetBstSublistMode{f}
\mciteSetBstMaxWidthForm{subitem}{(\alph{mcitesubitemcount})}
\mciteSetBstSublistLabelBeginEnd
  {\mcitemaxwidthsubitemform\space}
  {\relax}
  {\relax}

\end{mcitethebibliography}

\end{document}


\begin{figure}[H]
\begin{center}
\includegraphics[width=0.5\textwidth]{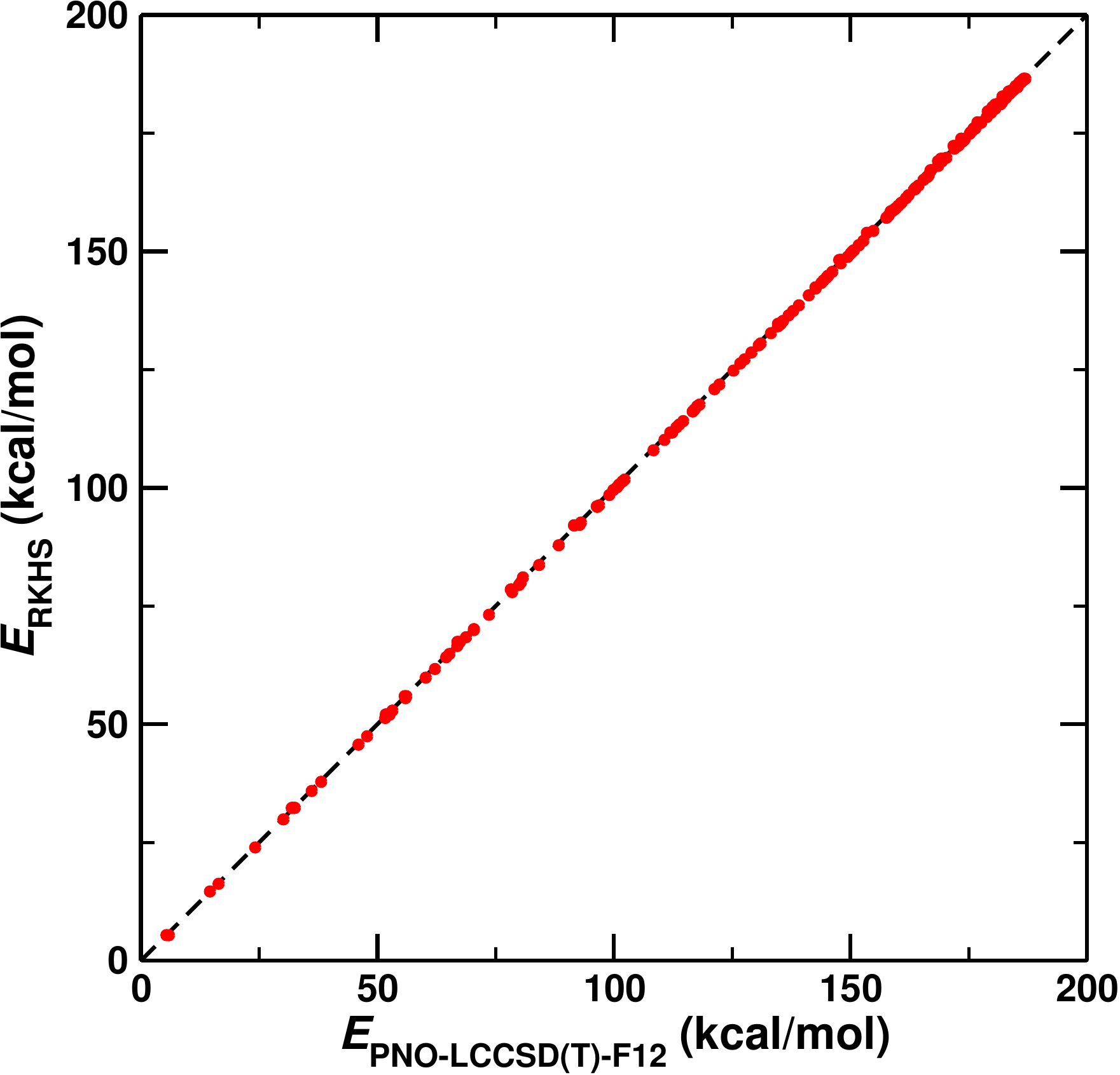}
\caption{Correlation between {\it ab initio} and RKHS interpolation
  for 230 randomly selected geometries. The $R^2 = 0.9999$ and the
  RMSD is 0.38 kcal/mol over a range of 180 kcal/mol.}
\label{sifig:offgrid}
\end{center}
\end{figure}

\begin{figure}[H]
\begin{center}
\includegraphics[width=0.7\textwidth]{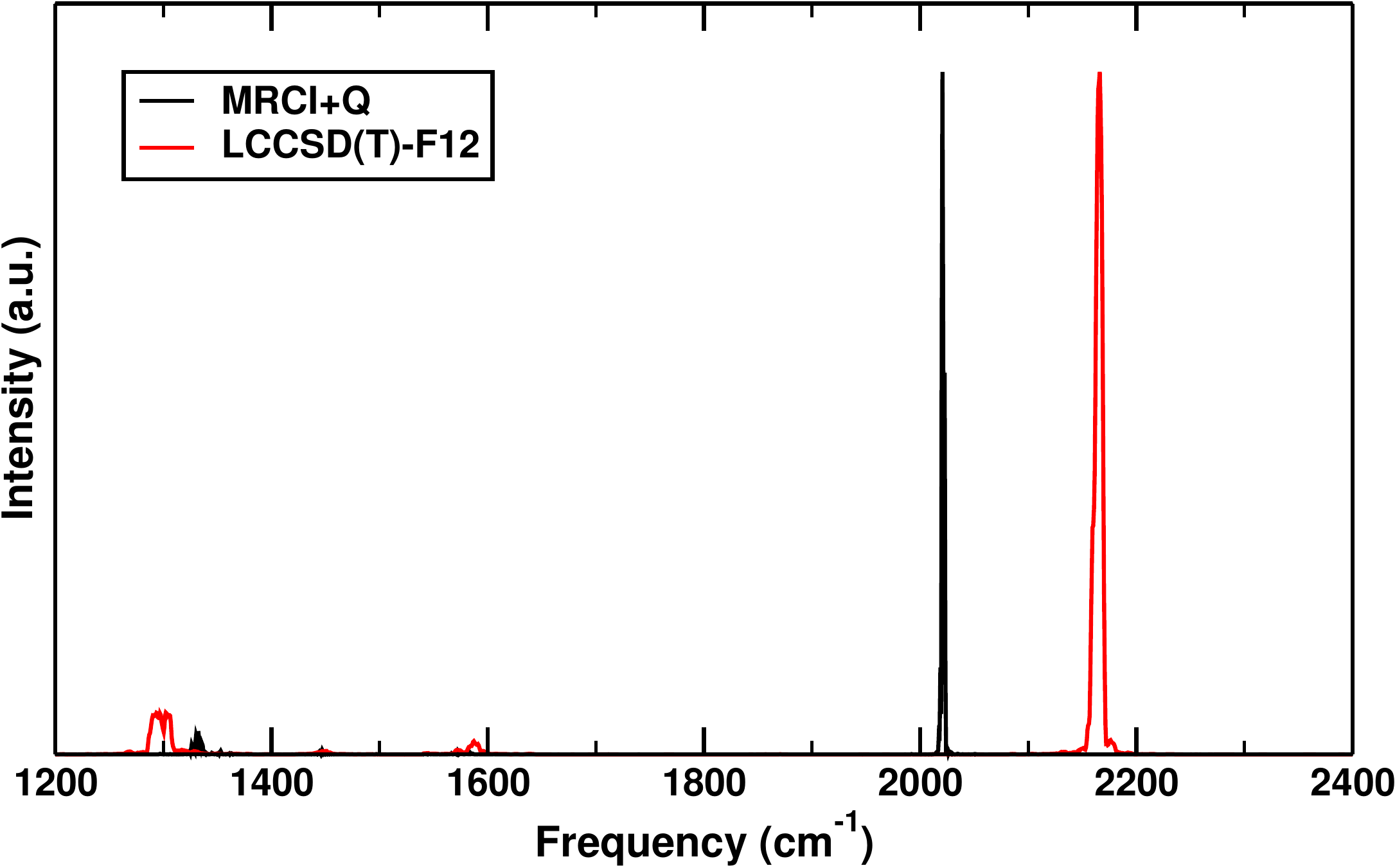}
\caption{Power spectrum based on the N1-N2 separation for AHA in the
  gas phase and from MRCI+Q and LCCSD(T)-F12 surface.}
\label{sifig:aha}
\end{center}
\end{figure}

\begin{figure}[H]
\centering
\includegraphics[width=0.4\textwidth,angle=-90]{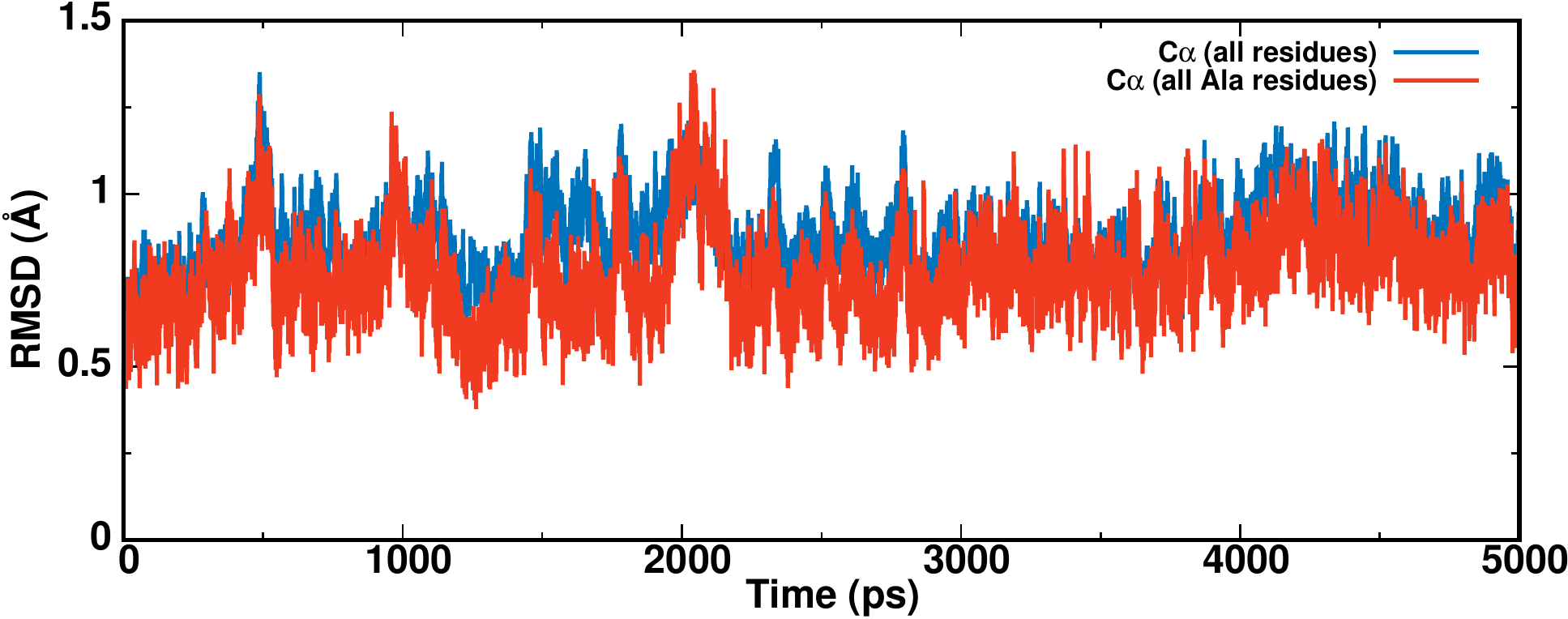}
\caption{The structural RMSD for the C$_{\alpha}$ atoms from all
  residues (blue) and for the 14 Ala residues (red) for WT Lysozyme.}
\label{sifig:rmsd_lys}
\end{figure}

\begin{figure}[H]
\begin{center}
\includegraphics[width=0.65\textwidth]{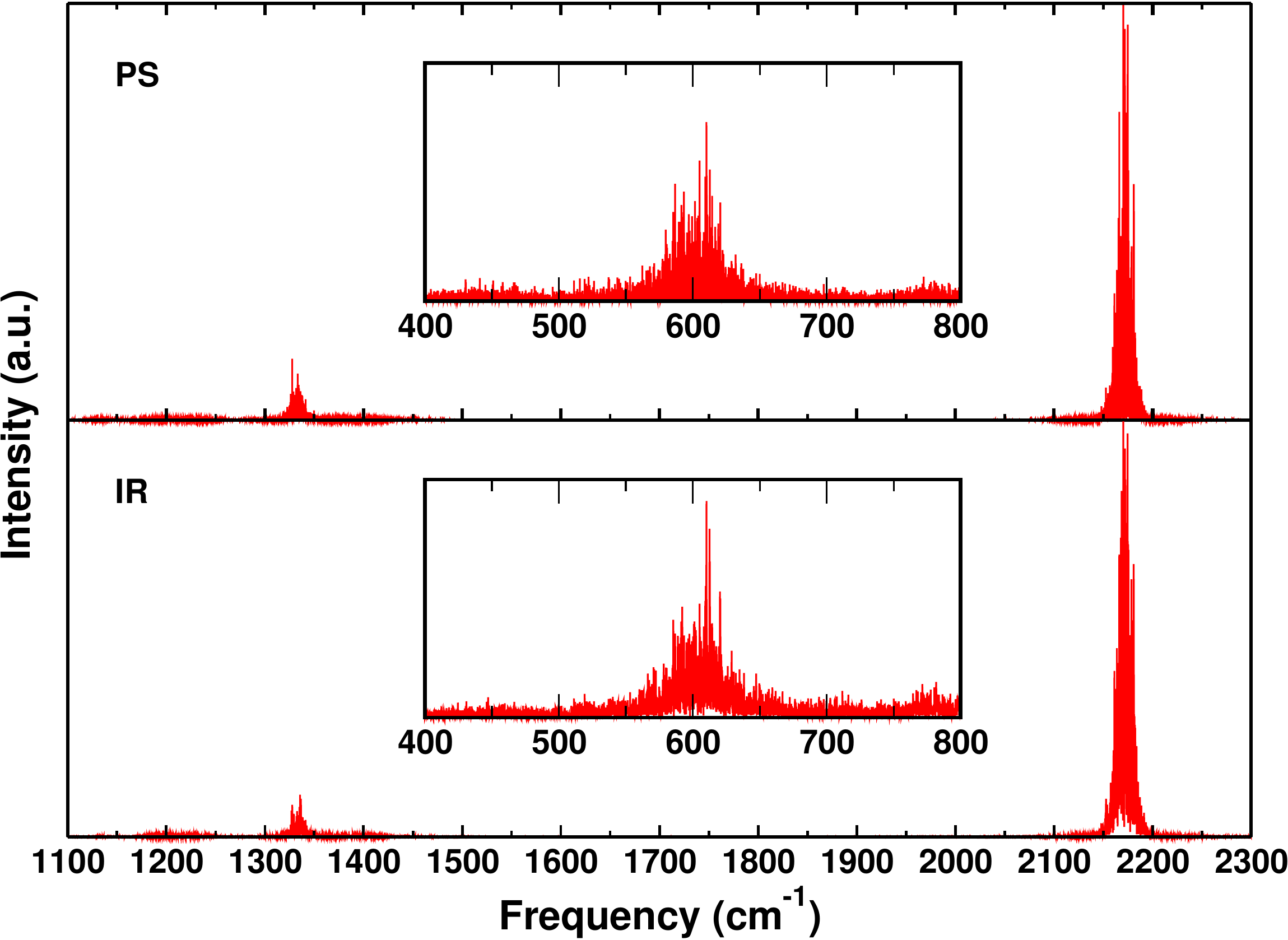}
\caption{Power (PS, top panel) and IR (bottom panel) spectrum for
  Ala47N$_3$. The power spectrum is based on N2-N3 bond
  displacement. The inset shows the bending mode of the Azide
  group. IR spectrum is calculated the Fourier transform of the
  molecular dipole moment autocorrelation function.}
\label{sifig:ps_irdip}
\end{center}
\end{figure}

\begin{figure}[H]
\begin{center}
  \includegraphics[width=0.5\textwidth,angle=-90]{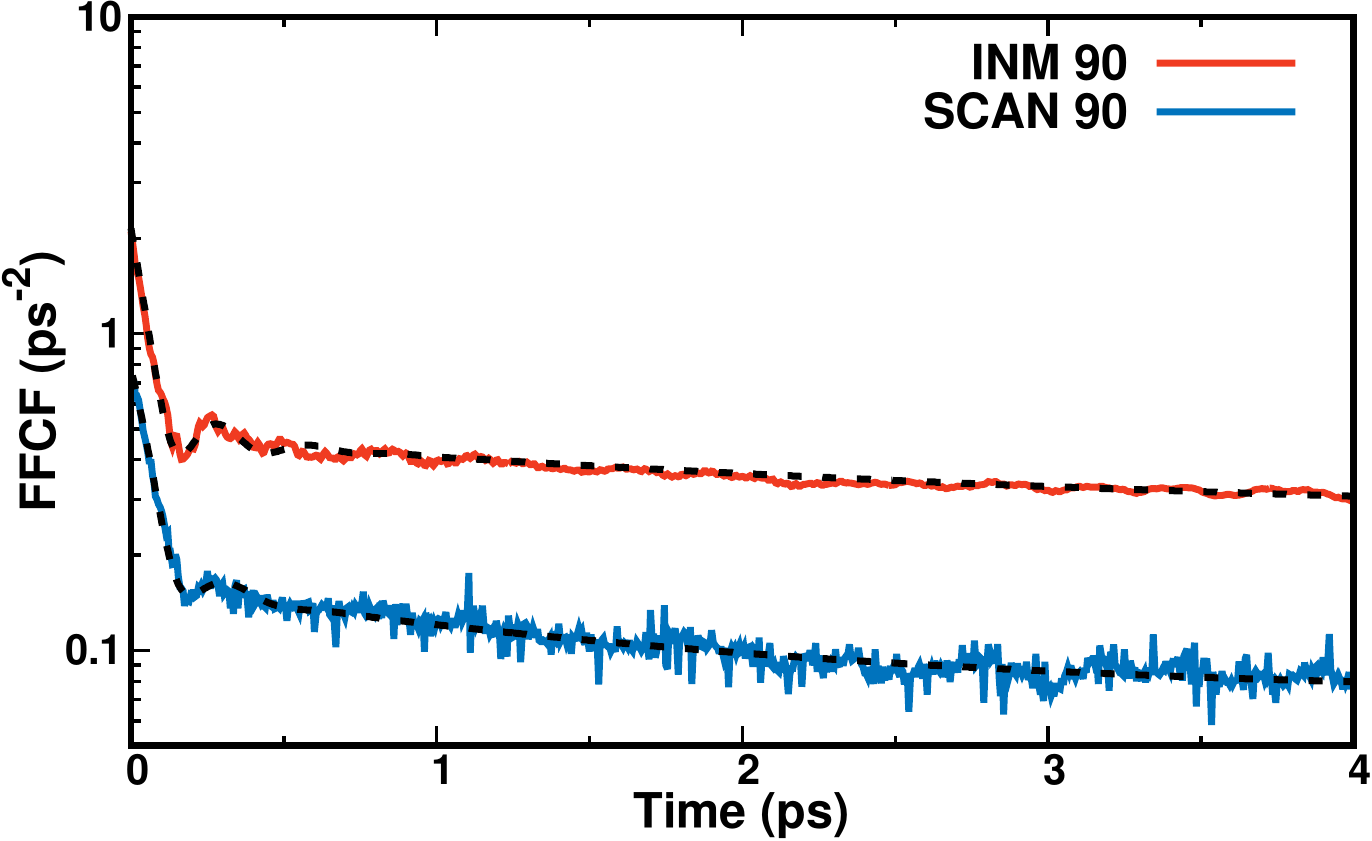}
\caption{The FFCF for Ala90 based on INM (red) and scan (blue)
  frequencies. The dashed lines show the corresponding fit to
  Eq. \ref{eq:ffcffit} with 3 time scales and the fitting parameters
  are as follows: INM: $a_{1}=0.29$, $\gamma=21.00$, $\tau_{1}=0.18$
  ps, $a_{2}=1.38$, $\tau_{2}=0.05$ ps, $a_{3}=0.22$, $\tau_{3}=2.70$
  ps, $\Delta_0=0.25$ and scan: $a_{1}=0.21$, $\gamma=17.50$,
  $\tau_{1}=0.11$ ps, $a_{2}=0.39$, $\tau_{2}=0.08$ ps, $a_{3}=0.09$,
  $\tau_{3}=1.62$ ps, $\Delta_0=0.07$. The comparison shows that the
  two different ways to determine the instantaneous frequency
  ($\omega(t)$ and $\nu(t)$, respectively) does not affect the overall
  appearance of the FFCF except for the magnitude of the asymptotic
  value $\Delta_0$.}
\label{sifig:fitscan}
\end{center}
\end{figure}

\begin{figure}[H]
\begin{center}
\includegraphics[width=0.38\textwidth,angle=-90]{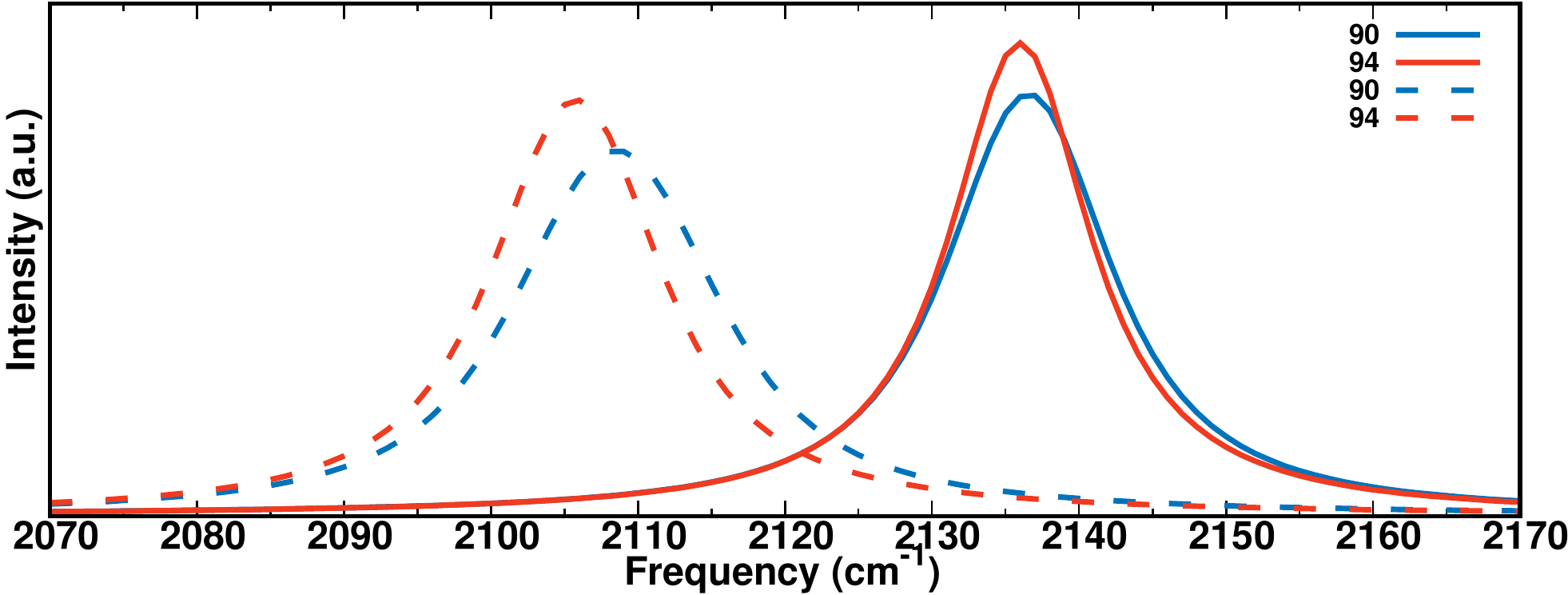}
\caption{1D IR spectra for Ala90N$_3$ and Ala94N$_3$ of lysozyme
  obtained from frequency calculations using ``scan'' (solid lines)
  and INM (dashed lines). Both analyses agree in that the frequency
  maximum for Ala90N$_3$ is to the blue of that for Ala94N$_3$ but the
  magnitude of the shift differs for the two approaches.}
\label{sifig:scan}
\end{center}
\end{figure}

\begin{figure}[H]
\begin{center}
\includegraphics[width=0.4\textwidth,angle=-90]{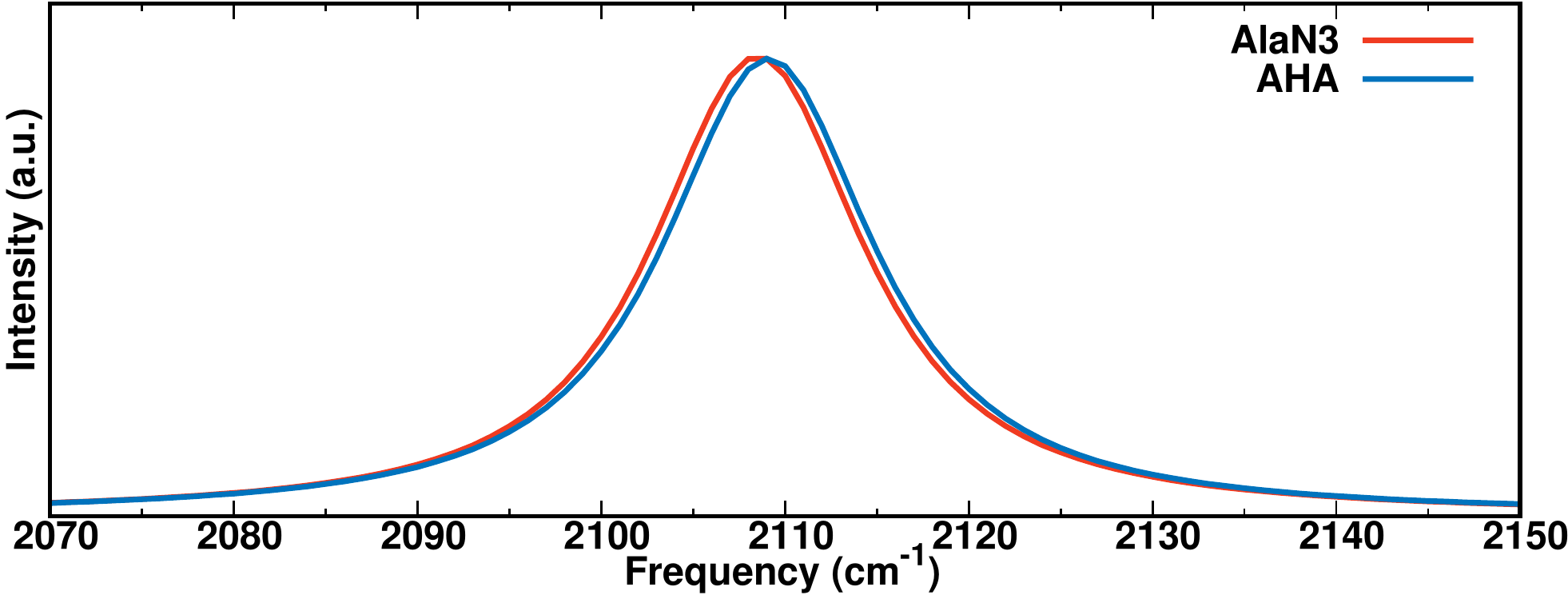}
\caption{Comparison between the 1D-IR spectra of AlaN$_3$ and AHA at
  position Ala47 in Lysozyme. For AHA, an additional CH$_2$ group is
  inserted before the N$_3^-$ label. The lineshapes for azidoalanine
  and azidohomoalanine are very similar. The position of the frequency
  maximum for the asymmetric stretch of the azide label differs by
  less than 1 cm$^{-1}$.}
\label{sifig:aha_alan3}
\end{center}
\end{figure}

\begin{figure}[H]
\begin{center}
\includegraphics[width=0.75\textwidth]{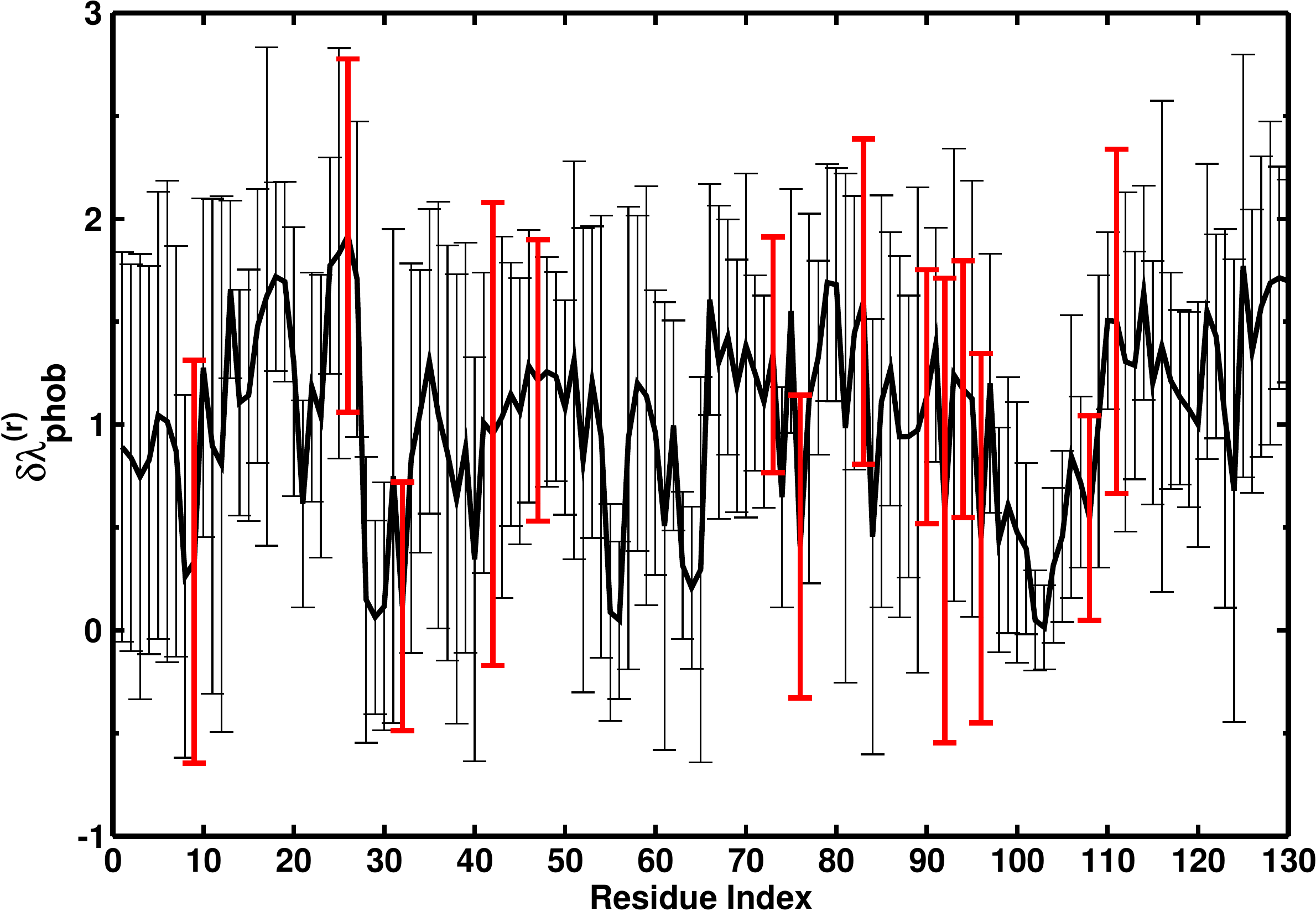}
\caption{Average local hydrophobicity together with fluctuations for
  WT Lysozyme (no N$_3^{-}$ labels attached) for the 5 ns
  simulation. LH for all residues in black and for Ala in red.}
\label{sifig:lh}
\end{center}
\end{figure}

\begin{figure}[H]
\centering
\includegraphics[width=0.4\textwidth,angle=-90]{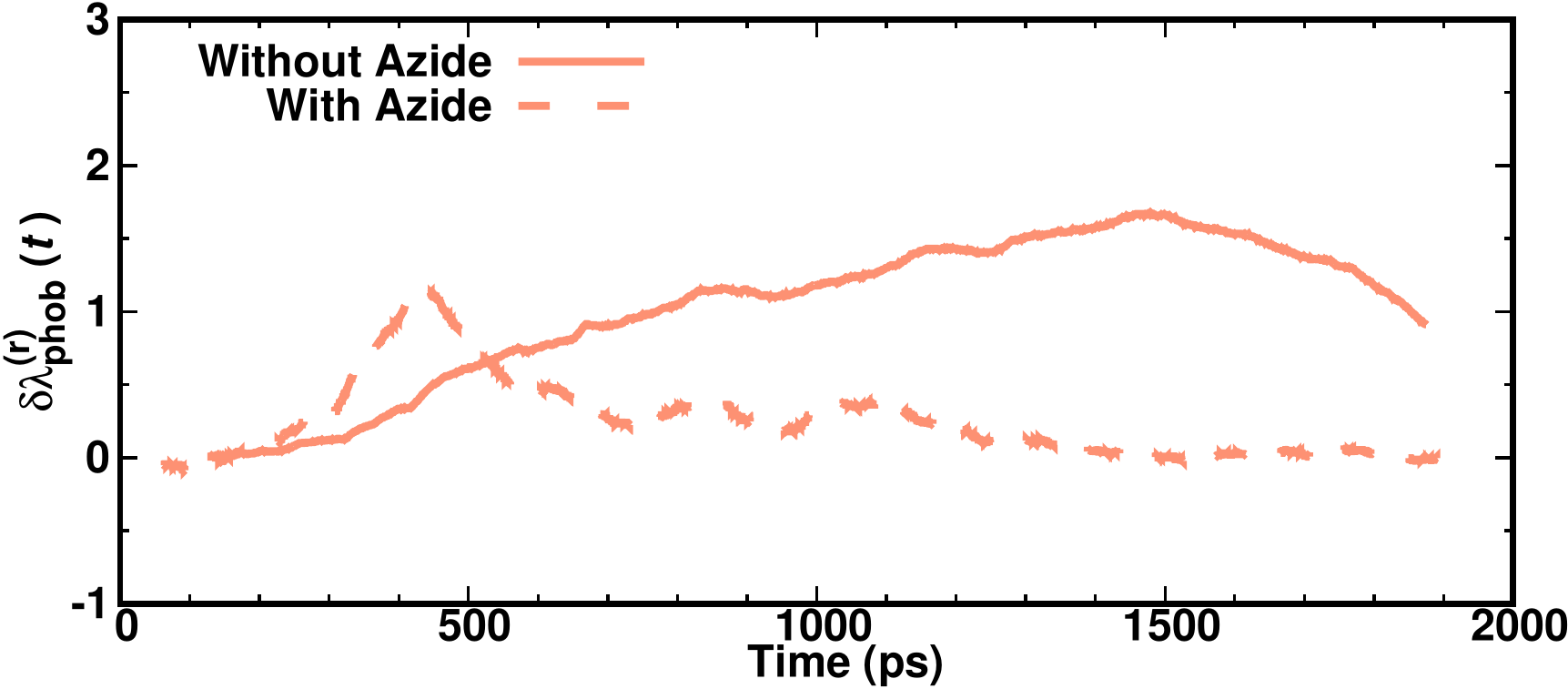}
\caption{Local hydrophobicity for residue 76 with and without azide
  group attached to Ala for 2 ns. The effect of attaching azide on the
  LH, and hence the hydration itself, is clearly visible.}
\label{sifig:hyd76}
\end{figure}